\documentclass[twocolumn,prb,preprintnumbers,amsmath,amssymb,superscriptaddress,showpacs]{revtex4-1}
\usepackage{graphicx}
\usepackage{setspace}

\usepackage{color}
\usepackage{ulem} 

\usepackage{amsmath,amsthm,amssymb}
\usepackage{mathrsfs}

\makeatletter
\def\Left#1#2\Right{\begingroup%
   \def\ts@r{\nulldelimiterspace=0pt \mathsurround=0pt}%
   \let\@hat=#1%
   \def\sht@im{#2}%
   \def\@t{{\mathchoice{\def\@fen{\displaystyle}\k@fel}%
          {\def\@fen{\textstyle}\k@fel}%
          {\def\@fen{\scriptstyle}\k@fel}%
          {\def\@fen{\scriptscriptstyle}\k@fel}}}%
   \def\g@rin{\ts@r\left\@hat\vphantom{\sht@im}\right.}%
   \def\k@fel{\setbox0=\hbox{$\@fen\g@rin$}\hbox{%
      $\@fen \kern.3875\wd0 \copy0 \kern-.3875\wd0%
      \llap{\copy0}\kern.3875\wd0$}}%
      \def\pt@h{\mathopen\@t}\pt@h\sht@im%
      \Right}%
\def\Right#1{\let\@hat=#1%
   \def\st@m{\mathclose\@t}%
   \st@m\endgroup}
\makeatother


\begin{document}
\title{Nonequilibrium steady states of electric-field driven Mott insulators}

\author{Yuta Murakami}
\affiliation{Department of Physics, University of Fribourg, 1700 Fribourg, Switzerland}
\author{Philipp Werner}
\affiliation{Department of Physics, University of Fribourg, 1700 Fribourg, Switzerland}
\date{\today}

\begin{abstract}
We present a systematic study of the nonequilibrium steady states (NESS) in Mott insulators driven by DC or AC electric fields, based on the Floquet dynamical mean-field theory. The results are analyzed using a generalized tunneling formula for the current, which is reminiscent of the Meir-Wingreen formula and provides insights into the relevant physical processes. In the DC case, the spectrum of the NESSs exhibits Wannier-Stark (WS) states associated with the lower and upper Hubbard bands. In addition, there emerge WS sidebands from many-body states. 
Using the tunneling formula, we demonstrate that the tunneling between these WS states leads to peaks or humps in the induced DC current.
In the AC case, we cover a wide parameter range of excitation frequencies and field strengths to clarify the crossover from field-induced tunneling behavior in the DC limit to nonequilibrium states dominated by multiphoton absorption in the AC limit.  
In the crossover regime, the single-particle spectrum is characterized by a coexistence of Floquet sidebands and WS peaks, and the current and double occupation exhibits a nontrivial dependence on the field strength. 
The tunneling formula works quantitatively well even in the AC case, and we use it to discuss the potential cooperation of tunneling and multi-photon processes in the crossover regime.
The tunneling formula and its simplified versions also provide physical insights into the high-harmonic generation in Mott insulators.
\end{abstract}


\maketitle

\section{Introduction}

Pump-probe spectroscopy has become a versatile tool for the investigation of nonequilibrium effects in correlated materials.\cite{Giannetti2016} In these experiments, the material is driven out of equilibrium by a strong laser pulse (pump), and its properties during or after the applied field pulse are measured by weaker probe pulses. 
Since the hopping times in typical correlated solids are of the order of femto seconds, THz field pulses may be regarded as quasi-static fields, while optical pulses induce inter-band transitions or so-called Floquet states with novel time-averaged properties. 

Mott insulators are strongly correlated materials, where nonequilibrium phase transitions induced by both types of excitations have been extensively studied. These systems would be metallic according to band theory, but are insulating in equilibrium because the charge motion is blocked by strong Coulomb interactions. 
Photo-induced phase transitions to a nonthermal metal state by 1.55 eV laser pulses have been demonstrated in organic materials,\cite{Okamoto2007,Wall2010} a nickel chain compound\cite{Iwai2003} and in cuprates,\cite{Okamoto2011} while a dielectric breakdown using (quasi-)static fields has recently been observed in Sr$_2$CuO$_2$,\cite{Taguchi2000dbMott} in VO$_2$\cite{Mayer2015dbMott} and an ET-based compound.\cite{Yamakawa2017dbMott}  
These complementary types of field-induced phase transitions have also been studied theoretically in one dimensional models using analytical methods, exact diagonalization and time-dependent density matrix renormalization group calculations, \cite{Takahashi2002,Oka2003,Matsueda2004,Oka2005,Yonemitsu2005,Oka2008,Takahashi2008,Oka2012,Zala2012,Meisner2010dcmott} 
in two dimensional models\cite{Takahashi2002,Shinjo2017}
and in higher dimensions using time-dependent Gutzwiller\cite{Mazza2015} and nonequilibrium dynamical mean field calculations.\cite{Eckstein2010c,Eckstein2011,Eckstein2013,Eckstein2013b,Werner2015,Murakami2017c,Francesco2017}
While these works have provided important insights into the nonequilibrium dynamics of Mott insulators in strong fields, such as the threshold behavior of the field-induced current in the dielectric breakdown case, or the energy distribution of photo-doped carriers after a resonant excitation, 
we still lack a systematic investigation of 
the general case where both the field amplitude and driving frequency are of the order of the characteristic energy scales of the system. 

Even though previous studies have mainly focused on transient dynamics and isolated interacting systems are expected to heat up to infinite temperature under continuous driving,\cite{Mori2016,Kuwahara2016} it should be noted that recent theoretical works have shown that long-lived quasi-steady states (prethermal states) can be rapidly reached, both in the case of near-resonant driving\cite{Murakami2017c,Francesco2017}
and in the static limit,\cite{Eckstein2010c,Eckstein2013b} 
 and that the characteristic properties of Floquet states are induced already by few-cycle pulses.\cite{Eckstein2017} 
In addition, realistic systems are coupled to an environment and the energy injected by the field can be  dissipated.  
 In this case, the system does not reach the infinite temperature state, but instead approaches a stable Floquet nonequilibrium steady state (NESS) under continuous driving.\cite{Aron2012dbMott} 
Thus, for a better understanding of strong field effects in Mott insulators driven by a pump pulse, it is useful to perform a systematic investigation of the properties of these NESSs.
So far, the NESSs of Mott insulating systems have been investigated in the DC limit\cite{Joura2008,Tsuji2008,Lee2014, Aron2012dbMott,Jiajun2015} and some AC regimes,\cite{Schmidt2002,Tsuji2008,Tsuji2009,Mikami2016,Murakami2017d} but the general features 
in the two-dimensional space of driving frequency and driving amplitude 
remain to be revealed. 

In this paper, we provide a systematic study of the NESS properties of the electric-field driven Mott insulating Hubbard model using a Floquet implementation of dynamical mean-field theory (Floquet DMFT).\cite{Tsuji2008,Tsuji2009,Aoki2013,Murakami2017}  
In order to stabilize a true NESS, this formalism involves a coupling to a  heat bath, which here consists of free fermions.  
While the parameters of the heat bath have an effect on the NESSs, we discuss qualitative features which are robust against the choice of bath parameters.
The Floquet DMFT is implemented with the non-crossing approximation (NCA) as an impurity solver, which is reliable in the strong-coupling regime.
Hence our analysis mainly focuses on systems with a large Coulomb interaction $U$, i.e. deep in the Mott regime.

In the DC limit, we show that our set-up reproduces the qualitative results of the previous studies on isolated systems, such as Wannier-Stark states and the associated resonances in the field-induced current.\cite{Eckstein2013b} Furthermore, we reveal additional WS states connected to many-body processes that are different from the usual Hubbard band resonances. 
In the AC case, we study a wide parameter range covering the DC limit and the weak-field AC limit and discuss the behavior of the spectral function, induced current and double occupation in the crossover regime which connects these two limits.
In order to obtain physical insights into the involved processes, we also introduce a generalized tunneling formula, which is applicable both to the DC and AC cases.
We demonstrate that this formula is also useful to investigate the high-harmonic generation (HHG) in Mott insulators.

The paper is organized as follows.
In Sec.~\ref{sec:formalism}, we explain the Floquet dynamical mean-field theory (FDMFT) implemented with the non-crossing approximation (NCA) as an impurity solver.\cite{Murakami2017d,Eckstein2010b} 
Furthermore, we introduce the generalized tunneling formula for the current in NESSs and its simplified versions.
In Sec.~\ref{sec:dc}, we discuss the properties of NESSs induced by DC fields, while  
in Sec.~\ref{sec:ac}, we analyze the NESSs induced by AC fields and discuss HHG based on the tunneling formula. 
Section~\ref{sec:conclusions} contains a summary and conclusions. 

\section{Formalism}\label{sec:formalism}
\subsection{Model and Method}
We consider the half-filled Hubbard model attached to a 
thermal bath and driven by an electric field,
\begin{align}
H(t)=&-\sum_{\langle i,j\rangle,\sigma} v_{ij}(t) c_{i,\sigma}^\dagger c_{j,\sigma}+U\sum_i n_{i\uparrow}n_{i\downarrow}\nonumber\\
&-\mu \sum_i n_i+H_{\rm bath}.
\end{align}
Here $c^\dagger_{i,\sigma}$ is the creation operator of an electron with spin $\sigma$ at the site $i$,
$U$ is the on-site Coulomb interaction, $v_{ij}$ is the hopping parameter and $\mu$ is the chemical potential.
The effect of the electric field is introduced by the Peierls substitution, $v_{ij}(t)=v_{ij}\exp\bigl(-iq\int^{{\bf r}_j}_{{\bf r}_i} d{\bf r} {\bf A}(t)\bigl)$,  
where the vector potential ${\bf A}(t)$ is related to the electric field by ${\bf E}(t)=-\partial_t {\bf A}(t)$. Here $q$ is the electron charge.
We consider a hyper-cubic lattice with lattice spacing $a$ in the limit of infinite spatial dimensions ($v=\frac{v^*}{2\sqrt{d}}$ with $d\rightarrow\infty$), 
which has a Gaussian density of states $\rho(\epsilon)=\frac{1}{\sqrt{\pi}v^*}\exp[-\epsilon^2/v^{*2}]$.
In the following, we set $q,a=1$ and use $v^*$ as the unit of energy. 
$H_{\rm bath}$ describes the coupling of the system to the environment (open system). We use a free electron bath (the B\"{u}ttiker model), 
with a finite band width $W_{\rm bath}$, whose retarded self-energy is $-{\rm Im}\Sigma^R_{\rm bath}(\omega)=\Gamma \sqrt{1-\left(\omega/W_{\rm bath}\right)^2}$.
The Keldysh and advanced components can be obtained from this through the fluctuation-dissipation theorem.\cite{Aoki2013, Murakami2017} 

 \begin{figure*}[t]
  \centering
    \hspace{-0.cm}
    \vspace{0.0cm}
   \includegraphics[width=140mm]{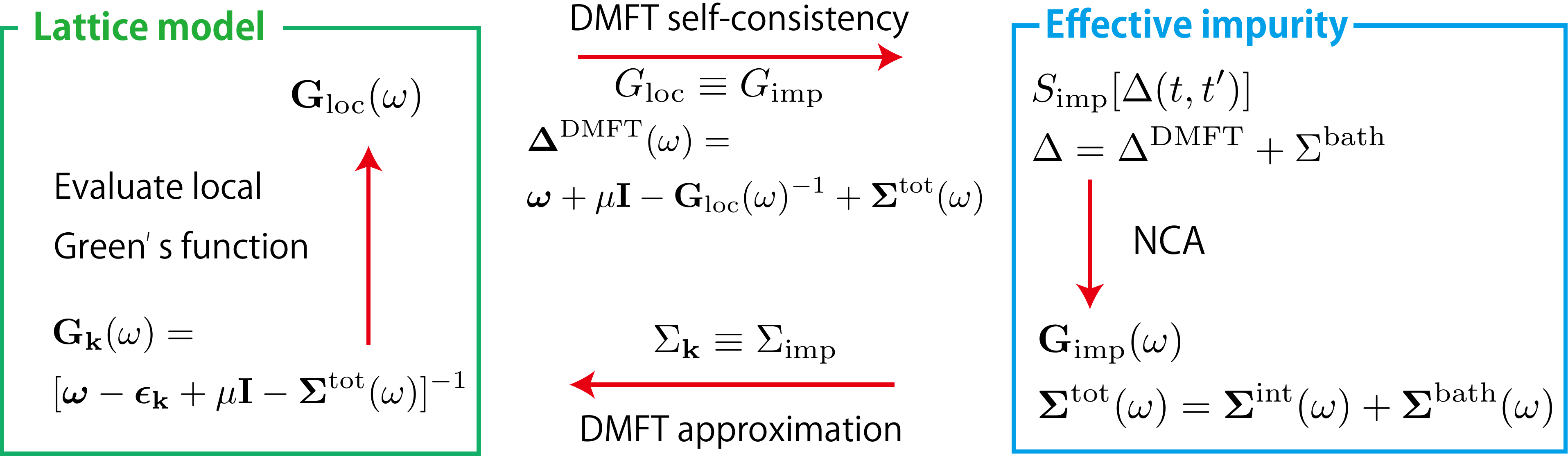} 
  \caption{FDMFT self-consistency loop. Boldified letters indicate the Floquet representation of the corresponding function. $S_{\rm imp}$ is the impurity action, 
  which is determined by the hybridization function. In the present case, the hybridization function consists of the effective bath which mimics the surrounding lattice ($\Delta^{\rm DMFT}$), and the attached electron bath ($\Sigma^{\rm bath}$). The problem is solved by NCA which yields the impurity Green's function and the self-energy ($\Sigma^{\rm tot}$) consisting of the contribution from the interaction 
  ($\Sigma^{\rm int}$) and that from the bath ($\Sigma^{\rm bath}$).
  Using this and $\epsilon_{\bf k}(t)=\epsilon ({\bf k}-{\bf A}(t))$ we obtain the lattice Green's function at each momentum, and by averaging over ${\bf k}$ the local Green's function. Through the impurity Dyson equation we update the hybridization function.}
  \label{fig:FDMFT_NCA_flow}
\end{figure*}

In the present study, we apply the DC and AC fields along the body diagonal, ${\bf A}(t)=A(t){\bf e}_0$ with ${\bf e}_0=(1,1,\cdots,1)$. 
In both cases, the system reaches a nonequilibrium steady state (NESS) in which the excitation by the field is balanced by the dissipation to the thermal bath. 
To be precise, in the DC case, the time-dependent vector potential is $qaA(t)=\Omega t$, corresponding to the field strength $E_0=-\frac{\Omega}{qa}$. In this gauge $H(t)$ oscillates with frequency $\Omega$, but the NESS exhibits no time-dependence of the physical observables.
On the other hand, for the AC field, we use $qaA(t)=A_0\sin \Omega t$ so that the field strength along a given axis is $E(t)=-\frac{A_0}{qa}\Omega \cos\Omega\equiv -E_0\cos\Omega t$.
In this case, the NESS is time-periodic with frequency $\Omega$.

Since both in the DC and AC field cases, the time-dependent Hamiltonian is time-periodic, 
we can use the Floquet dynamical mean-field theory (FDMFT) to determine the NESSs.\cite{Schmidt2002,Joura2008,Tsuji2008,Tsuji2009,Lee2014,Mikami2016,Murakami2017,Max2017,Murakami2017d} 
The main idea of DMFT is to evaluate the electron Green's function and self-energy by mapping the original lattice problem to an effective impurity problem.\cite{Georges1996}
An illustration of the DMFT self-consistency loop is presented in Fig.~\ref{fig:FDMFT_NCA_flow}.
The effective impurity problem is self-consistently determined such that the local lattice Green's function 
\begin{align}
G_{\rm loc}(t,t')= -i\langle \mathcal{T}_{\mathcal C} c_{i,\sigma} (t) c_{i,\sigma}^\dagger(t') \rangle
\end{align}
 is equal to the Green's function of the impurity site and the momentum-independent self-energy of the lattice is identical with the self-energy of the impurity problem. $\mathcal{T}_{\mathcal C}$ is the counter-ordering operator.
Here we assume a homogeneous system and a symmetry between spin up and down.
In the FDMFT we use the Keldysh contour, and the time-periodicity of the Green's function, $G_{\rm loc}(t+\mathcal{T},t'+\mathcal{T})=G_{\rm loc}(t,t')$. This allows us to switch to a 
Floquet representation of the Green's function and self-energy, and to express the Dyson equation in terms of matrix multiplications, 
see Refs.~[\onlinecite{Tsuji2008,Tsuji2009,Aoki2013,Murakami2017}] for further details.

The effect of the field enters through the time-periodic kinetic energy $\epsilon_{\bf k}(t)=\epsilon ({\bf k}-{\bf A}(t))$,
which appears in the lattice Dyson equation, see the left part of Fig.~\ref{fig:FDMFT_NCA_flow}. 
To implement the FDMFT, we use the non-crossing approximation (NCA) as an impurity solver. 
This lowest order self-consistent hybridization expansion is expected to produce qualitatively correct results deep in the Mott phase.\cite{Eckstein2010b}
The details of the implementation of the NCA impurity solver for nonequilibrium problems can be found in Ref.~\onlinecite{Eckstein2010b} and the supplemental material of Ref.~\onlinecite{Murakami2017d}.
In the present paper, we mainly focus on the large-gap Mott insulator ($U=8$). 

\subsection{Observables}
The observables used in this paper are defined as follows.
We are interested in the time-averaged local single-particle spectra,
\begin{align}
\bar{A}_{\rm loc}(\omega)=-\frac{1}{\pi} {\rm Im} \frac{1}{\mathcal{T}} \int_0^{\mathcal{T}} dt_a \int dt_r e^{i\omega t_r}G^R_{\rm loc}(t_r;t_a).
\end{align}
Here $R$ indicates the retarded part of the Green's function and $G^R_{\rm loc}(t_r;t_a)= G^R_{\rm loc}(t-t';\frac{t+t'}{2})=G^R_{\rm loc}(t,t')$. 
The time-averaged occupancy of the energy levels is given by
\begin{align}
\bar{N}_{\rm loc}(\omega)=\frac{1}{2\pi} {\rm Im} \frac{1}{\mathcal{T}} \int_0^{\mathcal{T}} dt_a \int dt_r e^{i\omega t_r}G^<_{\rm loc}(t_r;t_a),
\end{align}
where $<$ indicates the lesser part of the Green's function.
The distribution function is defined as ${\bar f}(\omega)=\bar{N}_{\rm loc}(\omega)/\bar{A}_{\rm loc}(\omega)$.
In the case of a DC field we do not need to take time averages and we just write $A_{\rm loc}$,  $N_{\rm loc}$ and $f$.

The doublon density, the current and the kinetic energy are measured as 
\begin{align}
d(t)&=\langle n_{i,\uparrow}(t)n_{i,\downarrow}(t)\rangle,\\
j(t)&=\frac{q}{N}\sum_{{\bf k},\sigma} {\bf e}_0\cdot {\bf v}_{\bf k}(t)  \langle c_{{\bf k},\sigma}^\dagger(t) c_{{\bf k},\sigma}(t)\rangle,\\
E_{\rm kin}(t)&=\frac{1}{N}\sum_{{\bf k},\sigma}\epsilon_{\bf k}(t) \langle c_{{\bf k},\sigma}^\dagger(t) c_{{\bf k},\sigma}(t)\rangle.
\end{align}
Here $N$ is the system size, ${\bf k}$ is the lattice momentum, $c^{\dagger}_{{\bf k},\sigma}=\frac{1}{\sqrt{N}} \sum_i e^{i{\bf k}\cdot{\bf r}_i} c^\dagger_{i,\sigma}$ and ${\bf v}_{\bf k}(t)=\partial_{\bf k} \epsilon ({\bf k}-{\bf A}(t))$.
Let us also introduce another expression for the current in terms of $G_{\bf k}(t,t')=-i\langle \mathcal{T}_{\mathcal{C}} c_{{\bf k},\sigma}(t)c^{\dagger}_{{\bf k},\sigma}(t')\rangle $, namely
\begin{align}\label{eq:MW_eqlike}
j(t)&=i \frac{q}{N}\sum_\alpha \sum_{{\bf k},\sigma} \epsilon_{\bf k}(t) \bigl[\partial_{k_\alpha}G_{\bf k}^<(t,t)\bigl]\nonumber\\
&=i\frac{q}{N} \sum_{{\bf k},\sigma}  \epsilon_{\bf k}(t) \int^t_{-\infty} d\bar{t} \Bigl[\sum_\alpha \partial_{k_\alpha}\epsilon_{\bf k}(t)\Bigl]\nonumber\\
&\;\;\;\;\;\;\;\times[G_{\bf k}^>(t,\bar{t})G_{\bf k}^<(\bar{t},t)-G_{\bf k}^<(t,\bar{t})G_{\bf k}^>(\bar{t},t)].
\end{align}
To derive this expression, we use partial integration and the formula
\begin{align}
\partial_{k_{\alpha}} G_{\bf k}(t,t')=\int_{\mathcal{C}} d\bar{t} G_{\bf k}(t,\bar{t}) [\partial_{k_\alpha} \epsilon_{\bf k}(\bar{t}) ] G_{\bf k}(\bar{t},t'),
\end{align}
where $\int_{\mathcal{C}}$ is the contour integral along the Keldysh contour. The latter expression can be derived by taking the derivative with respect to ${\bf k}$ of the Dyson equation and assuming a momentum-independent self-energy ($G_{\bf k}^{-1}=G_{0,{\bf k}}^{-1}-\Sigma$).

In the present case of a hyper-cubic lattice and an electric field pointing in the body-diagonal direction, the momentum (${\bf k}=(k_1,k_2,\cdots,k_d)$) dependence of the Green's function 
can be parameterized by $\epsilon=-2v\sum_{i=1}^d \cos(k_i)$ and $\bar{\epsilon}=-2v\sum_{i=1}^d \sin(k_i)$, i.e., $G_{\bf k}(t,t')=G_{\epsilon,\bar{\epsilon}}(t,t')$. 
Using this one can express the current as 
\begin{align}\label{eq:j_exact}
j(t)&=-\frac{q}{2} \int d\epsilon d\bar{\epsilon} \rho(\epsilon,\bar{\epsilon})\int^t_{-\infty}F(\epsilon,\bar{\epsilon},t,\bar{t})\times\nonumber\\
&[G_{\epsilon,\bar{\epsilon}}^>(t,\bar{t})G_{\epsilon,\bar{\epsilon}}^<(\bar{t},t)-G_{\epsilon,\bar{\epsilon}}^<(t,\bar{t})G_{\epsilon,\bar{\epsilon}}^>(\bar{t},t)],
\end{align}
with
\begin{align}
F(\epsilon,\bar{\epsilon},t,\bar{t})=&\big[(\epsilon^2+\bar{\epsilon}^2)e^{i(A(t)-A(\bar{t}))}-\text{H.c.}\\
&-(\epsilon-i\bar{\epsilon})^2e^{i(A(t)+A(\bar{t}))}-\text{H.c.}\big].\nonumber
\end{align}
Here $\rho(\epsilon,\bar{\epsilon})=\frac{1}{\pi v^{*2}}\exp[-\frac{\epsilon^2+\bar{\epsilon}^2}{v^{*2}}]$ is the joint density of states.

\subsection{Tunneling formula for the current in NESSs}

Here we introduce a tunneling formula for the current in NESSs and simplified versions of it.
A more detailed derivation can be found in Appendix A.
For the derivation, we choose one direction in the hyper-cubic lattice (let's say $x$) and 
regard the system as a stack of $(d-1)$-dimensional slabs, which are aligned in the $x$ direction.
The bias is applied to the $x$ direction and each slab is connected to the neighboring slab by $l^{d-1}$ 
(with $l\rightarrow\infty$) tunneling junctions.
The Hamiltonian can now be expressed as $\hat{H}(t)=\hat{H}_{\perp}(t)+\hat{V}_{x}(t)$, where $\hat{V}_{x}(t)$
describes the transfer integrals along the $x$ direction (junctions between slabs), 
and $\hat{H}_{\perp}(t)$ the $(d-1)$-dimensional slabs.
In order to prepare the NESS of the full system, we start from the Floquet steady state of decoupled slabs ($V_x=0$), and switch on $V_x$ adiabatically.
We consider the effect of $V_x$ in the linear response and evaluate the current along the $x$ direction (through the junctions between the slabs) 
considering that the steady states of the $(d-1)$-dimensional slabs can be approximated by that 
of the full $d$-dimensional bulk when $d$ is large.
Then we obtain the following expression of the current per site
\begin{align}\label{eq:j_tun_mom}
&j_{\rm tun,mom}(t)=-qv^{*2} \int d\epsilon d\bar{\epsilon} \rho(\epsilon,\bar{\epsilon})
\int^t_{-\infty} d\bar{t}\\
& {\rm Re}\Bigl[ G^<_{\epsilon,\bar{\epsilon}}(\bar{t},t) G_{\epsilon,\bar{\epsilon}}^>(t,\bar{t})-G^>_{\epsilon,\bar{\epsilon}}(\bar{t},t) G_{\epsilon,\bar{\epsilon}}^<(t,\bar{t})\Bigl]e^{-i\int^t_{\bar{t}}dt' E(t')}.\nonumber
\end{align}
We note that this expression includes processes where an electron goes through a certain junction and returns to the original slab through a different junction, see Appendix A.
It is also interesting to point out that by approximating $\epsilon^2=\bar{\epsilon}^2\simeq v^{*2}/{2}=\int d\epsilon d\bar{\epsilon}  \rho(\epsilon,\bar{\epsilon})\epsilon^2$ and $\epsilon\bar{\epsilon}\simeq 0 =\int d\epsilon d\bar{\epsilon}  \rho(\epsilon,\bar{\epsilon})\epsilon \bar{\epsilon}$, Eq.~(\ref{eq:j_exact}) becomes Eq.~(\ref{eq:j_tun_mom}).

By only considering the contribution to the current at a certain junction by electrons that went  through the same junction,
we obtain the generalized tunneling formula,
\begin{widetext}
\begin{align}
j_{\rm tun}(t)=-qv^{*2}&{\rm Re}\Bigl[\int_{-\infty}^t d\bar{t} \{G_{\rm loc}^<(\bar{t},t)G_{\rm loc}^>(t,\bar{t})-G_{\rm loc}^>(\bar{t},t)G_{\rm loc}^<(t,\bar{t})\}e^{-i\int^t_{\bar{t}}dt' E(t')}\Bigl]. \label{eq:j_tun_tot}
\end{align}
\end{widetext}
This formula is applicable both to the DC field and the AC field case.
In the following sections, we will show that the tunneling formula Eq.~(\ref{eq:j_tun_tot}) works quantitatively very well 
in a wide parameter range, both in the DC and AC excitation regime.  
We will also show that the formula is useful to study the HHG in Mott insulators.
These analyses indicate that, at least in the parameter range studied in the paper, the neglected processes are not important.
We also note that the expressions Eqs.~(\ref{eq:j_exact}), (\ref{eq:j_tun_mom}), (\ref{eq:j_tun_tot}) are reminiscent of 
the Meir-Wingreen formula that is often used in quantum dot and transport problems.\cite{Meir1992} 

In the special case of NESSs driven by a DC field, using $G_{\rm loc}^{<>}(t,t')=G_{\rm loc}^{<>}(t-t')$ and $E(t)=E_0$, 
Eq.~(\ref{eq:j_tun_tot}) can be expressed as 
\small
\begin{align}\label{eq:j_tun_dc}
&j_{\rm tun}(E_0)=qv^{*2}\pi \int^\infty_\infty d\omega A_{\rm loc}(\omega) A_{\rm loc}(\omega+E_0)\\
&\times [f_{\rm loc}(\omega)(1-f_{\rm loc}(\omega+E_0))-f_{\rm loc}(\omega+E_0)(1-f_{\rm loc}(\omega))].\nonumber
\end{align}
\normalsize
The same formula for DC driven systems has been derived in Ref.~\onlinecite{Lee2014} based on an expansion in $v^*$.

Next we introduce some simplified versions of Eq.~(\ref{eq:j_tun_tot}), which are inspired by naive expectations.
In the AC case, $G_{\rm loc}^{<>}(t_r;t_{\rm av})$ in general depends on $t_{\rm av}$ and  $G_{\rm loc}^{<>}(t,t')\ne G_{\rm loc}^{<>}(t-t')$. 
In other words, there are contributions from the off diagonal terms of $G_{\rm loc}$ in the Floquet representation and Eq.~(\ref{eq:j_tun_tot}) cannot be 
expressed only with $t_{\rm av}$-averaged quantities.
Still it is an interesting question how well the current is approximated by $t_{\rm av}$-averaged quantities.
To investigate this issue, we introduce a simplified version of Eq.~(\ref{eq:j_tun_tot}) by replacing $G^{<>}_{\rm loc}(t,t')$ by $\bar{G}^{<>}_{\rm loc}(t-t')$, i.e., we approximate the current as
\small
\begin{align}\label{eq:j_tun_av}
&j_{\rm tun,av}(t)=-qv^{*2}\times\\
&{\rm Re}\Bigl[\int^{\infty}_0 d\bar{t} \{\bar{G}_{\rm loc}^<(-\bar{t})\bar{G}_{\rm loc}^>(\bar{t})-\bar{G}_{\rm loc}^>(-\bar{t})\bar{G}_{\rm loc}^<(\bar{t})\}e^{-i\int^t_{t-\bar{t}}dt' E(t')}\Bigl]. \nonumber
\end{align}
\normalsize

Furthermore, one may consider a scenario where the current originates from tunneling induced by a quasi-static field between the Floquet states of two slabs .
This picture can be tested by assuming that $E(t')$ is slowly (adiabatically) changing so that for the current at time $t$ the field $E(t')$ in the integral can be replaced by $E(t)$. 
This leads to the approximation 
\small
\begin{align} \label{eq:j_tun_adi}
&j_{\rm tun,adi}(t)=-qv^{*2}\times\\
&{\rm Re}\Bigl[\int_{-\infty}^t d\bar{t} \{G_{\rm loc}^<(\bar{t},t)G_{\rm loc}^>(t,\bar{t})-G_{\rm loc}^>(\bar{t},t)G_{\rm loc}^<(t,\bar{t})\}e^{-i(t-\bar{t}) E(t)}\Bigl]\nonumber.
\end{align}
\normalsize

Finally, one may combine the above two assumptions (time-average of the Green's function and the adiabaticity of the field) to obtain
\small
\begin{align}\label{eq:j_tun_av_adi}
&j_{\rm tun,av,adi}(t)=-qv^{*2}\times\\
&{\rm Re}\Bigl[\int^{\infty}_0 d\bar{t} \{\bar{G}_{\rm loc}^<(-\bar{t})\bar{G}_{\rm loc}^>(\bar{t})-\bar{G}_{\rm loc}^>(-\bar{t})\bar{G}_{\rm loc}^<(\bar{t})\}e^{-i\bar{t}E(t)}\Bigl] \nonumber\\
&=qv^{*2}\pi \int^\infty_\infty d\omega \bar{A}_{\rm loc}(\omega) \bar{A}_{\rm loc}(\omega+E(t))\times\nonumber\\
& [\bar{f}_{\rm loc}(\omega)(1-\bar{f}_{\rm loc}(\omega+E(t)))-\bar{f}_{\rm loc}(\omega+E(t))(1-\bar{f}_{\rm loc}(\omega))].\nonumber
\end{align}
\normalsize
This formula captures the tunneling induced by the quasi-static field between the Floquet bands (photo-dressed states) 
appearing in the time-averaged spectrum.

Finally, let us emphasize that Eq.~(\ref{eq:j_tun_tot}) is not an exact formula.
One can show this by considering the linear response against $E(t)$.
By expanding Eq.~(\ref{eq:j_tun_tot}) in terms of $E(t)$, we find that the real part of the optical conductivity 
is given by
\begin{align}
{\rm Re} \sigma(\Omega)=\frac{q v^{*2}\pi}{\Omega} \int d\omega &A_{\rm loc,eq}(\omega)A_{\rm loc,eq}(\omega+\Omega)\nonumber\\ 
&\times[f_{\rm eq}(\omega)-f_{\rm eq}(\omega+\Omega)].
\end{align}
Here we used the fact that the inversion symmetry leads to the absence of a linear component in $E(t)$ in $G_{\rm loc}$ and the subscript ``${\rm eq}$" indicates equilibrium quantities.
On the other hand, the correct expression involves convolutions at each momentum, with momentum-dependent single particle Green's functions.\cite{Georges1996}
One can show that Eq.~(\ref{eq:j_tun_mom}) provides the exact expression of the linear optical conductivity. 

\section{DC field driven steady states}\label{sec:dc}

First we discuss the effects of DC fields. 
When the applied field is strong enough, doublons and holons are created by tunneling  processes  
and the Mott insulating phase becomes unstable. 
This phenomenon is called dielectric break-down and the time evolution after the switch-on of the field has been studied with exact diagonalization, time-dependent density matrix renormalization group and DMFT calculations.\cite{Oka2003,Oka2005,Eckstein2010c} 
In the present study, we are interested in the properties of the NESSs that are reached after the transient dynamics in the presence of $H_\text{bath}$.

 \begin{figure}[t]
  \centering
    \hspace{-0.cm}
    \vspace{0.0cm}
   \includegraphics{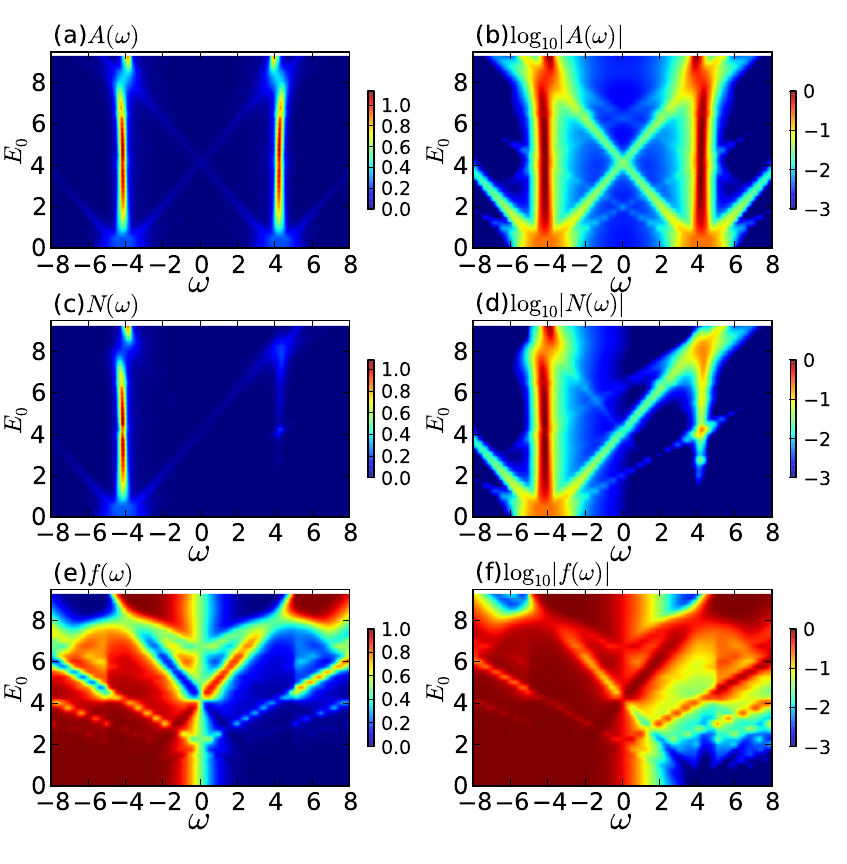} 
  \caption{(a)(b) The local spectral function $A(\omega)$, (c)(d) the occupation function $N(\omega)$ and  (e)(f) the distribution function $f(\omega)$ of the Hubbard model with DC field.  The parameters are  $U=8.0,\beta=2.0,\Gamma=0.06,W_{\rm bath}=5.0$.}
  \label{fig:fig1}
\end{figure}

\subsection{Spectral function and Occupation}
First we discuss the spectral function and the occupation in the NESS.
In Fig.~\ref{fig:fig1}, we show the local spectral function (panels (a)(b)), their occupancy (panels (c)(d)) and the distribution function (panels (e)(f)). 
Without the field, the lower and upper Hubbard bands are located around $\omega=\pm U/2$, and their band width is about $2v^*$.
When the field is applied, Wannier-Stark states emerge from both the lower and upper Hubbard bands ($\omega_{{\rm ws},\pm,m}\simeq \pm U/2 +m E_0$).
For $E_0\gtrsim U/2$, other peaks that scale as $\omega_{{\rm ws2},\pm,\mp2}=\pm\frac{3}{2} U\mp 2E$ start to become prominent, see Fig.~\ref{fig:fig1}(b).
The side-band $\omega_{{\rm ws2},+,-2}=\frac{3}{2} U-2E_0$ represents a process where the insertion of one electron at a given site leads to the creation of a doublon and the simultaneous creation of another doublon-holon pair.
This doublon/holon creation process costs an interaction energy of $\frac{3}{2}U$ (if the chemical potential contribution is subtracted).
If the separation between the additional doublon and holon in the field direction is two lattice spacings, there is a potential energy gain of $2E_0$.
Without this energy gain from the field, the considered process is strongly suppressed and only produces a weak shoulder in the local spectral function at $\omega=\pm \frac{3}{2}U$, but our simulations show that it becomes relevant for $E_0\gtrsim U/2$. 
There is also a very weak feature that follows $\omega_{{\rm ws2},\pm,\mp1}=\pm\frac{3}{2} U\mp E_0$ (only visible in a second derivative plot of the local spectral function). It corresponds to the process of creation of one doublon (holon) at a given site and the simultaneous creation (annihilation) of another doublon-holon pair with a separation of one lattice spacing in the field direction.  
The satellites at $\omega=\pm \frac{3}{2}U$ and the side-bands emerging from them are the result of electronic correlations.  
In a semiconductor model, these structures at $\omega_{{\rm ws2},\pm,\mp2}=\pm\frac{3}{2} U\mp 2E_0$ are absent, see Appendix~\ref{app:semi}. 
By increasing the field we observe a crossing of the main Hubbard bands ($\omega_{0,\pm}\simeq \pm \frac{U}{2}$) and the first Wannier-Stark states ($\omega_{{\rm ws},\pm,\mp1}\simeq \pm \frac{U}{2}\mp E_0$)
around $E_0\simeq U$. 
The spectrum in Fig.~\ref{fig:fig1}(a)(b) shows that there occurs a hybridization between the main band and the Wannier-Stark state,
 which increases the splitting between the main bands ($|\omega_{0,+}-\omega_{0,-}|$) for $E_0\lesssim U$ and deceases it for $E_0\gtrsim U$.

We note that our results from FDMFT implemented with the NCA solver predict different properties of the spectral function around $E_0=U/2$ from those obtained in Ref.~\onlinecite{Lee2014}. In the latter work, FDMFT was implemented with an  iterated perturbation theory (IPT) impurity solver and a Markovian quantum master equation was used to describe the effect of a bosonic thermal bath. This calculation produced Wannier-Stark states emerging from the field-induced mid-gap states at $E_0=U/2$, but did not exhibit the peaks at $\omega=\pm\frac{3}{2} U\mp 2E_0$.  
The origin of these discrepancies remains unclear. 

Now we turn to the occupation ($N(\omega)$) and the distribution function ($f(\omega)$). 
The Wannier-Stark bands emerging from 
the lower Hubbard band and its satellites 
are well occupied and prominently visible in $N(\omega)$. There are also maxima along these bands in $f(\omega)$. 
On the other hand, the Wannier-Stark bands emerging from $\omega>0$ are less occupied and 
there are valleys (local minima) along these lines in $f(\omega)$.
Still a clear occupation of $\omega_{{\rm ws},+,-1}$ appears for $E_0\gtrsim \tfrac{U}{2}$, which  
corresponds to the annihilation of an already existing doublon by the removal of an electron.
For $E_0\gtrsim \tfrac{U}{2}$, there already exist many doublons in the NESS because of the activated tunneling processes.

\subsection{Current and double occupation}
\begin{figure}[t]
  \centering
    \hspace{-0.cm}
    \vspace{0.0cm}
   \includegraphics{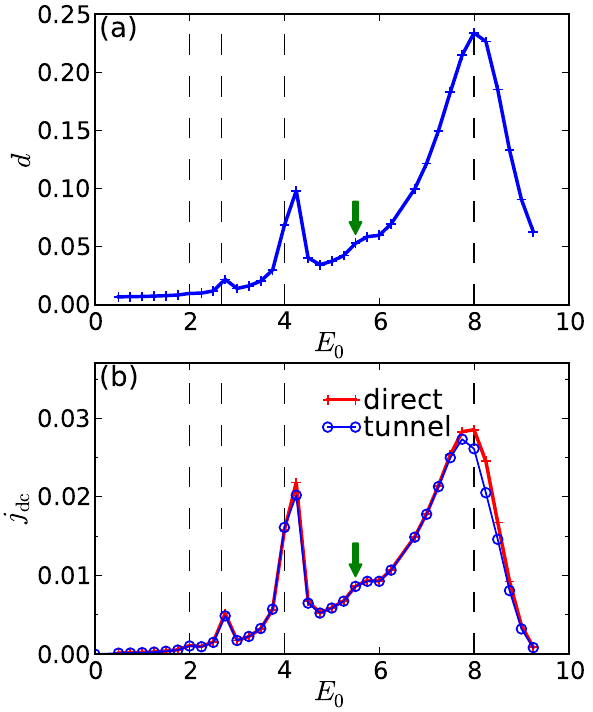} 
  \caption{(a)(b) Field-strength dependence of the double occupancy ($d$) 
  and the current ($j_{\rm dc}$) of the Hubbard model with DC field. 
  In panel (b), the measured current is shown with red cross marks, while the current estimated by the tunneling formula, Eq.~(\ref{eq:j_tun_dc}), is shown with blue circles.
  Vertical lines indicate $E_0=U$, $U/2$, $U/3$, and $U/4$, and the arrows indicate humps at $E_0\simeq5.5\simeq \frac{2}{3}U$. Here, $U=8.0$, $\beta=2.0$, $\Gamma=0.06$, $W_{\rm bath}=5.0$.}
  \label{fig:fig2}
\end{figure}

Now we move on to the double occupancy ($d$) and the induced DC current ($j_{\rm dc}$), see Fig.~\ref{fig:fig2}. 
Both in $d$ and $j_{\rm dc}$, one can observe clear peaks at $E_0=U$, $U/2$, $U/3$ and a small peak at $E_0=U/4$.
At $E_0=U/m$, the excess potential energy from tunneling of an electron by $m$ sites exactly compensates the energy required for the creation of a doublon-holon pair. Hence the doublon-holon pairs with a separation of $m$ lattice spacings are resonantly created and enhance the double occupancy and the current.
We also note that there is a hump around $E_0\simeq5.5\simeq \frac{2}{3}U$ both in $d$ and $j_{\rm dc}$. 
This feature corresponds to the simultaneous creation of two doublon-holon pairs, one with a separation of two lattice spacings and the other with a separation of one lattice spacing in the direction of the field.  
Again, such a resonance should be absent in the non-interacting case, which we have confirmed by analyzing a semiconductor model,
see Appendix~\ref{app:semi}.

These interpretations of the peak/hump structures in $d$ and $j_\text{dc}$ can be 
 corroborated by considering the tunneling formula for the DC current, Eq.~(\ref{eq:j_tun_dc}),
 which associates these structures with Wannier-Stark side-bands and main Hubbard bands in the spectrum. 
The results of Eq.~(\ref{eq:j_tun_dc}) match very well the directly evaluated current, see Fig.~\ref{fig:fig2}(b).
Using the tunneling formula, the $E_0=U/3$ peak, for example, can be explained as follows.
For this field strength, the spectral function exhibits clear peaks at i) $-\frac{U}{2}=\omega_{{\rm ws},-,0}$, 
ii) $-\frac{U}{2}+\frac{U}{3}=\omega_{{\rm ws},-,1}=\omega_{{\rm ws},+,-2}$, iii)$\frac{U}{2}-\frac{U}{3}=\omega_{{\rm ws},-,2}=\omega_{{\rm ws},+,-1}$
and  iv) $\frac{U}{2}=\omega_{{\rm ws},+,0}$, which are equally spaced by $\frac{U}{3}$.
According to Eq.~(\ref{eq:j_tun_dc}), these peaks are the origin of the peak in $j_{\rm tun}$ at $\Omega=U/3$.
Even though in ii) and iii) two different WS states are mixed, we have to remember that $\omega_{{\rm ws},-}$ tends to be more occupied than $\omega_{{\rm ws},+}$.
Therefore, the main contribution should come from a) $\omega_{{\rm ws},-,0}\rightarrow \omega_{{\rm ws},+,-2}$, b) $\omega_{{\rm ws},-,1}\rightarrow \omega_{{\rm ws},+,-1}$
and c) $\omega_{{\rm ws},-,2}\rightarrow \omega_{{\rm ws},+,0}$. 
Altogether, these processes describe tunneling over three sites, because $\omega_{{\rm ws},\pm,m}$ corresponds to a doublon or holon mainly located at the $|m|$-th 
neighbor of the site on which we measure the local Green's function. 
The tunneling formula also correctly reproduces the hump at $E_0\simeq \frac{2}{3}U$. 
This can be attributed to the peaks at  i) $-\frac{U}{2}=\omega_{{\rm ws},-,0}$, 
ii) $-\frac{U}{2}+\frac{U}{3}=\omega_{{\rm ws}2,-,2}=\omega_{{\rm ws},+,-1}$, iii)$\frac{U}{2}-\frac{U}{3}=\omega_{{\rm ws},-,1}=\omega_{{\rm ws}2,+,-2}$
and  iv) $\frac{U}{2}=\omega_{{\rm ws},+,0}$, which are equally spaced by $\frac{U}{3}$.
The tunneling from i) to iii) and ii) to iv) is the origin of the hump at $E_0\simeq \frac{2}{3}U$, which is associated with the creation of two doublon-holon pairs, as mentioned above.

 \begin{figure*}[t]
  \centering
    \hspace{-0.cm}
    \vspace{0.0cm}
   \includegraphics[width=140mm]{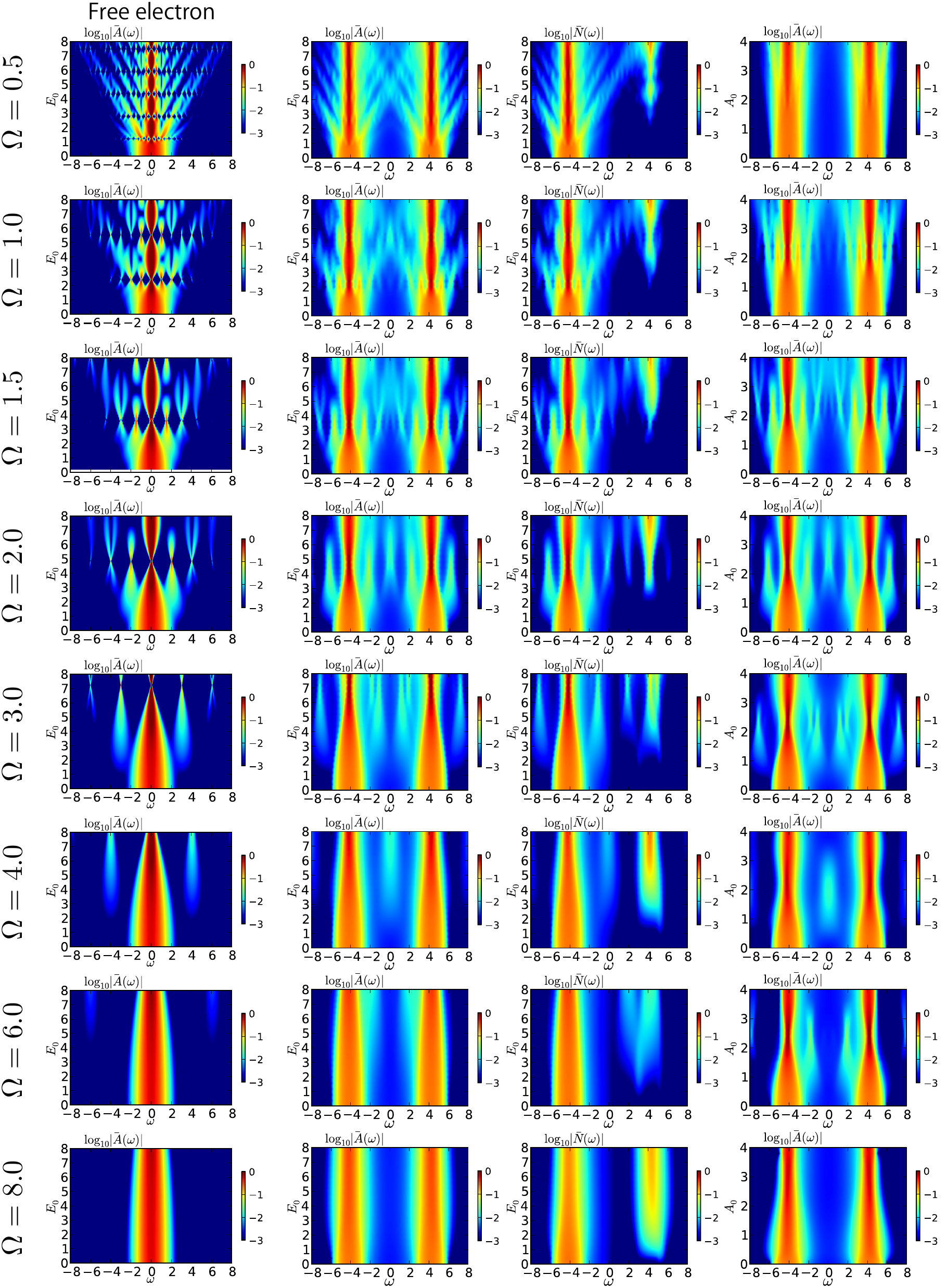} 
  \caption{Log scale plots of the local spectral function $\bar{A}(\omega)$ and occupation function $\bar{N}(\omega)$ of the Hubbard model driven by AC fields with indicated frequencies $\Omega$ and amplitudes $E_0$. 
  Here, $U=8.0,\beta=2.0,\Gamma=0.06,W_{\rm bath}=5.0$. }
  \label{fig:fig3s}
\end{figure*}

 \begin{figure}[t]
  \centering
    \hspace{-0.cm}
    \vspace{0.0cm}
   \includegraphics[width=60mm]{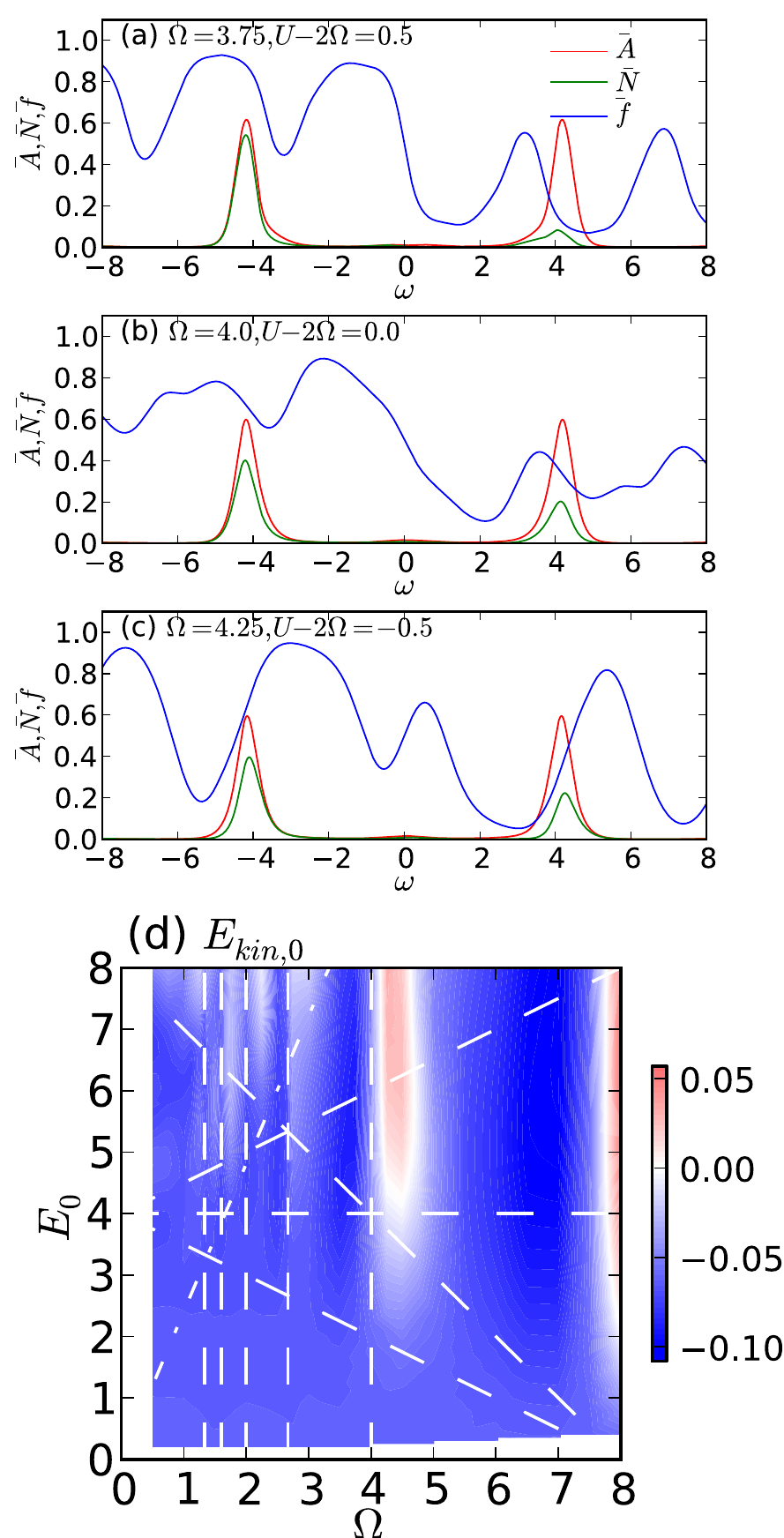} 
  \caption{ (a-c) Time-averaged local spectrum, occupation and distribution function around $\Omega=U/2$. The field strength is  $E_0/\Omega=1.841$. 
  (d)Time-averaged kinetic energy in the plane of $E_0$ and $\Omega$.  Here, $U=8.0,\beta=2.0,\Gamma=0.06,W_{\rm bath}=5.0$. }
  \label{fig_AC_spectrum2}
\end{figure}

\section{AC driven steady states}\label{sec:ac}
Next we discuss the effects of AC fields by considering various sets of parameters for 
the excitation frequency $\Omega$ and the field strength $E_0$.
In particular, we are interested in the properties of NESSs 
in the crossover regime ($v^*\lesssim \Omega,E_0$), where the 
intuition gained from the DC limit and the weak-field AC limit may not be applicable anymore and the two pictures
are expected to get mixed up.

\subsection{Spectral function and Occupation}
 \begin{figure*}[t]
  \centering
    \hspace{-0.cm}
    \vspace{0.0cm}
   \includegraphics[width=160mm]{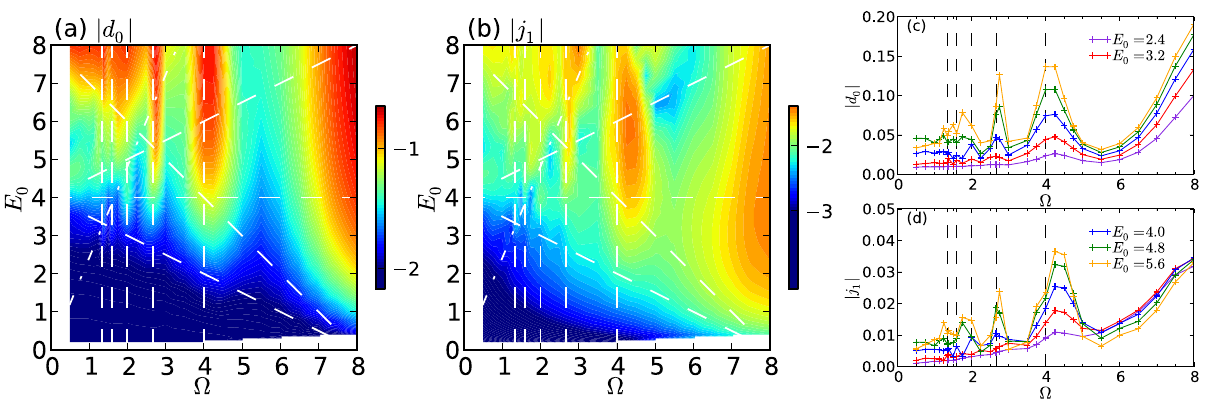} 
  \caption{(a,b) Log scale plot of the static components of the double occupancy (time-averaged, $|d_{0}|$) and the 1st order harmonic component of the current ($|j_1|$) in the plane of the field strength $E_0$ and the excitation frequency $\Omega$. 
  (c)(d) Cuts for fixed $E_0$.  Here, $U=8.0,\beta=2.0,\Gamma=0.06,W_{\rm bath}=5.0$. }
  \label{fig:fig4}
\end{figure*}
In Fig.~\ref{fig:fig3s}, we summarize the time-averaged local spectral functions $\bar{A}_{\rm loc}(\omega)$ and time-averaged occupancies $\bar{N}_{\rm loc}(\omega)$ 
for $\Omega=0.5,1.0,1.5,2.0,3.0,4.0,6.0,8.0\leq U$ and $E_0\leq U$.
Here we note that $\Omega=0.5,1.0,2.0,4.0,8.0$ satisfy the resonant condition $n\Omega=U$.
In the spectrum, when $\Omega$ is small ($\lesssim 1.0$), one can still identify side peaks around the Hubbard bands that scale linearly with $E_0$, see $\Omega=0.5$.
These correspond to the Wannier-Stark states in the DC limit.
As we increase the excitation frequency, we enter a crossover regime. 
There, in addition to the Wannier-Stark ladders, one starts to observe clear features 
at $\pm \frac{U}{2}+m\Omega$, which are reminiscent of the Floquet side bands of the Hubbard bands expected in the high-frequency regime, see for example $\Omega=1.0$ and $\Omega=1.5$. 
As we further increase the excitation frequency, the Wannier-Stark-like features become less prominent and one can only observe the Floquet side bands of the Hubbard bands 
at $\pm \frac{U}{2}+m\Omega$.
In general, the width of the main bands is reduced in the presence of the field, compared to the equilibrium value, and it changes as a function of the field strength. 
This shows that under a strong field the injected charge carriers 
get localized.\cite{Tsuji2008,Freericks2008FK,Iwai2014,Ono2017} 
In particular, the minimum width is observed when $J_0(E_0/\Omega)=0$ as in the free electron case (see below),
and the Floquet side bands become also very prominent here. 
To clearly reveal this dependence, 
we plot 
$\bar{A}_{\rm loc}(\omega)$
as a function of the strength of the vector field, 
$A_0=E_0/\Omega$, 
see the right row of Fig.~\ref{fig:fig3s}.
Here, we find a minimum around $A_0=2.4$, which is around the first zero crossing of the Bessel function $J_0(A_0)$.
We note that the renormalization and hence the minimum is less prominent for $\Omega=4.0,8.0$.
This appears to be because the hybridization of the Floquet side bands of the lower 
and upper Hubbard band widens the band. The same should happen at $\Omega=2.0$ but the weights of the Floquet side bands are small compared to the other cases, so that this effect is not prominent.

It is instructive to compare the spectra of the driven Mott insulators with those of driven free electron systems.\cite{Tsuji2008} 
In the present case of the hyper-cubic lattice and the electric field pointing in the body-diagonal direction, 
the free electron spectral function is 
\begin{align}
&\bar{A}_{\epsilon,\bar{\epsilon}}(\omega)=\sum_m \delta(\omega-m\Omega-\epsilon J_0(A_0))\nonumber\\
\times&\Bigl|\int^{2\pi}_0\frac{dx}{2\pi} e^{imx}\exp[-\frac{i}{\Omega} \int^x_0 dz \epsilon[\cos(A_0\sin z)-J_0(A_0)]\nonumber\\
&+\bar{\epsilon}\sin(A_0\sin z)]\Bigl|^2,
\end{align}
where the momentum information is again encoded in $\epsilon=-2v\sum_{i=1}^d \cos(k_i)$ and $\bar{\epsilon}=-2v\sum_{i=1}^d \sin(k_i)$.
From this, one can see that at each momentum the spectrum shows peaks at $\epsilon J_0(A_0)+m\Omega$.
The time averaged local spectrum can be expressed as 
$\bar{A}_{\rm loc}(\omega)=\int d\epsilon d\bar{\epsilon} \rho(\epsilon,\bar{\epsilon}) \bar{A}_{\epsilon,\bar{\epsilon}}(\omega)$,
where $\rho(\epsilon,\bar{\epsilon})=\frac{1}{\pi v^{*2}}\exp[-\frac{\epsilon^2+\bar{\epsilon}^2}{v^{*2}}]$ is the joint density of states.
In the left row of Fig.~\ref{fig:fig3s}, we show the time-averaged local spectrum for the free electrons. 
One can easily recognize the similarity between the free electron spectrum and the upper/lower Hubbard bands of the Mott insulator under the periodic driving.
Namely, for small $\Omega$ compared to the band width $\simeq 2 v^*$, one can identify structures resembling the WS ladders,
while for higher frequencies one can identify clear Floquet side bands around $\omega=n\Omega$.
When $J_0(A_0)=0$, the position of the peaks is at $m\Omega$ regardless of the momentum and this leads to clear Floquet side bands.

As in the DC case, the side peaks originating from the lower Hubbard bands tend to be more occupied compared to those originating from the upper Hubbard bands. 
The occupation of the upper Hubbard bands becomes prominent around $E_0=U/2$ for small excitation frequencies $\Omega$.
This indicates that in this regime the tunneling picture of the DC limit still works and the creation of doublon-hole pairs on the next nearest neighbor site is activated at $E_0=U/2$.
At higher frequencies, the main upper Hubbard band starts to get occupied for smaller value of $E_0$, which indicates that in this regime, the doublons are not created by tunneling but rather by multi-photon processes.

We now discuss the similarities and differences between the NESSs and the Floquet prethermal states of 
isolated systems, which were recently discussed in Ref.~\onlinecite{Murakami2017c}.
In Fig.~\ref{fig_AC_spectrum2}(a-c), we show 
spectral functions, occupation functions and distribution functions
 for $E_0/\Omega=1.841$ around $\Omega=U/2$,
where 
$x=1.841$ corresponds to the first crossing of the 0th and 2nd Bessel function, $J_0(x)=J_2(x)$.
We note that when $\Omega=U/2$ and $J_0(E_0/\Omega)=J_2(E_0/\Omega)$, the leading-order effective Hamiltonian obtained from the high-frequency expansion becomes the free Hamiltonian.\cite{Bukov2016}
In the previous study, we found that the Floquet prethermal state is characterized by i) a periodic distribution function $\bar f(\omega)$, i.e. $\bar f(\omega)\simeq \bar f(\omega+\Omega)$ and ii) an inverted population around $\Omega=U/2$ for $\Omega>U/2$. The observation i) can be well explained by regarding the Floquet prethermal state as a thermal state of the effective Hamiltonian. 
Figure~\ref{fig_AC_spectrum2} shows that such a periodicity in $\bar f(\omega)$ is not observed in the NESS, even though there is some reminiscent structure.
From this, it seems unlikely that the NESS can be described as an equilibrium state of 
some effective Hamiltonian derived from a high-frequency expansion.
Still, we can observe a population inversion around $\Omega=U/2$ and $\Omega>U/2$, 
similar to Ref.~\onlinecite{Murakami2017c} to the inversion around $\Omega=U$ and $\Omega>U$ in a periodically driven Falicov-Kimball model.\cite{Tsuji2009}
This inverted population results in a positive time-averaged kinetic energy $E_{\rm kin,0}$, as is shown in Fig.~\ref{fig_AC_spectrum2}(d).
We note that in equilibrium $E_{\rm kin,0}$ should take a negative value.
One can also observe a tendency of $E_{\rm kin,0}$  becoming positive around higher-order resonances $\Omega=U/n$.

\subsection{Current and double occupation}

Now we move on to the behavior of the induced current $j(t)=\sum_n e^{in\Omega} j_n$ and the double occupation $d(t)=\sum_n e^{in\Omega} d_{n}$.
These quantities oscillate periodically and they include the components of higher harmonics in terms of $\Omega$.
Because of the inversion symmetry of the system, $j(t)$ only includes odd harmonics, while $d(t)$ has even harmonics.
In this section, we mainly focus on the lowest two components, namely, $j_1$ for the current and $d_{0}$ for the double occupancy.
We discuss  the higher order components at the end of this section and also in the following section, which focuses on high-harmonic generation.

The behavior of $d_{0}$ and $j_1$ in the DC limit and weak-field AC limit can be well understood in terms of tunneling and multi-photon absorption, respectively,
but the values of these quantities at finite $\Omega$ and $E_0$ are difficult to predict.   
From the behavior in these limits, one may expect resonances at $E_0=U/n$ and $\Omega=U/n$ to extend into the regions of finite frequency and significant field strength.
Somewhere between these two limits, there occurs a crossover from the nonlinear transport regime (DC-like response) to the nonlinear optics regime (AC-like response). 
The corresponding crossover line, known as the {\it Keldysh line},\cite{Keldysh1965} has been studied in 1D Mott insulators.\cite{Oka2012}
This crossover-line should scale as $E_0\propto \Omega$ with a slope that is determined by the characteristic doublon-holon separation. 
In addition one could naively expect an interplay between tunneling and multi-photon absorption processes in the crossover regime, for example field-induced tunneling between Floquet side-bands of the time-averaged spectral function.
If such processes were relevant, one should detect noticeable changes in $d_{0}$ and $j_1$ along certain lines in the $E_0$-$\Omega$ plane.  For example, the line $U=E_0+\Omega$ would delimit the region in which doublon-holon creation by the combination of tunneling plus single photon absorption becomes relevant, while the line $U=2E_0+\Omega$ corresponds to two tunnelings and a single photon absorption.

 \begin{figure}[t]
  \centering
    \hspace{-0.cm}
    \vspace{0.0cm}
   \includegraphics[width=86mm]{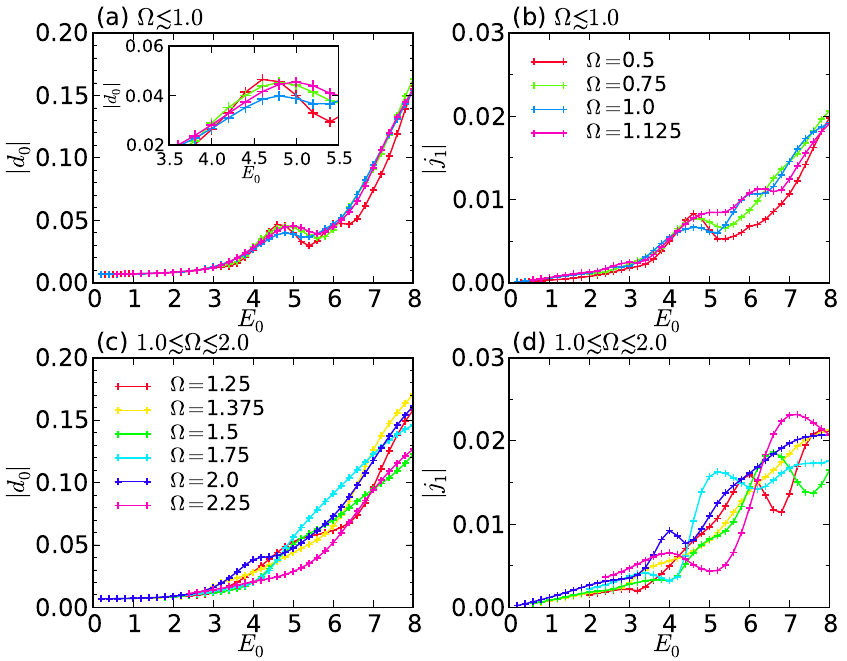} 
  \caption{$E_0$-dependence of $d_{0}$ and $j_1$ in the small-$\Omega$ regime. 
  Here, $U=8.0,\beta=2.0,\Gamma=0.06,W_{\rm bath}=5.0$.
  }
  \label{fig:fig6}
\end{figure}

In Fig.~\ref{fig:fig4}(a)(b) we plot the numerical results of our Floquet-DMFT (NCA) simulations in the space of $\Omega$ and $E_0$, with the naively expected crossovers or resonances indicated by white dashed lines. We do not show the Keldysh line, since DMFT does not provide a good estimate for the slope of this line, but we indicate the line $J_0(E/\Omega)=0$, which corresponds to complete localization (first zero crossing of the Bessel function) in the band renormalization picture. 
The actual behavior is obviously very complex and a detailed understanding of all the features in these plots is beyond the scope of this study.
One can easily identify peaks along the $\Omega=U/n$ lines in both $d_{0}$ and $j_1$
with clear resonances up to $n=6$, which are clearly visible around $E_0=4$. 
We plot curves for fixed field strength near $E_0=U/2=4$ as a function of $\Omega$ in Fig.~\ref{fig:fig4}(c)(d).
The general tendency for $E_0\lesssim U/2$, that higher order peaks become more prominent with larger $E_0$, is consistent with the expectation in the perturbative regime that  such processes are proportional to $|E_0|^{2n}$ for $d_{0}$ and $|E_0|^{2n-1}$ for $j_{1}$.
By increasing the field strength beyond $E_0 \approx U/2$, some higher-order
resonances disappear or emerge at off resonant conditions. 
These shifts in the peak positions may originate from a strong mixing of tunneling and multi-photon processes.

 \begin{figure}[t]
  \centering
    \hspace{-0.cm}
    \vspace{0.0cm}
   \includegraphics[width=86mm]{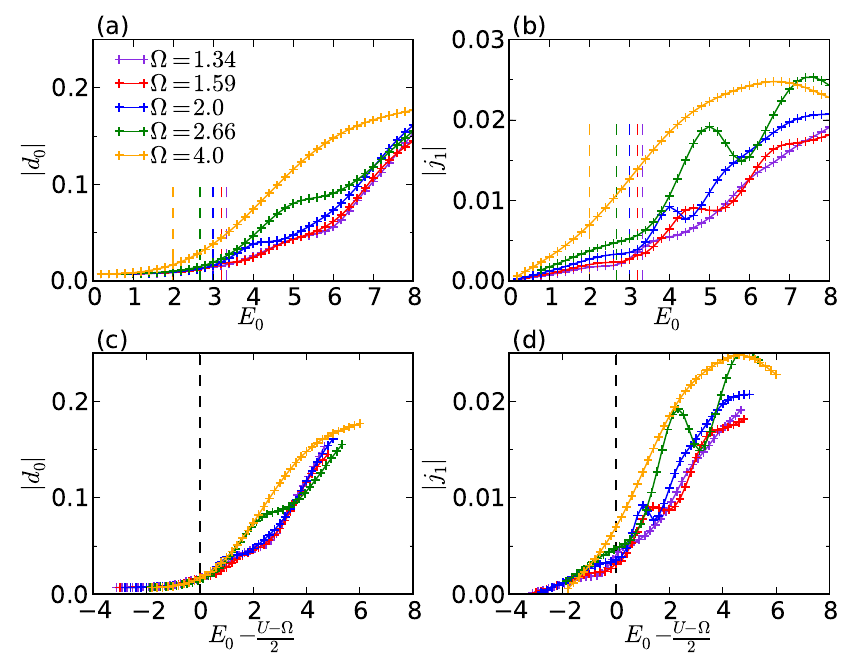} 
  \caption{$E_0$-dependence of $d_{0}$ and $j_1$ near the resonances $\Omega=U/n$. Panels (a)(b) show the
  original data, and panels (c)(d) the results with the horizontal axis shifted by the ``threshold field" $E_{0,\text{line}}=(U-\Omega)/2$.
  Vertical lines indicate $E_{0,\text{line}}$ for the parameters specified by the colors.
  Here, $U=8.0,\beta=2.0,\Gamma=0.06,W_{\rm bath}=5.0$.}
  \label{fig:fig_U_2E_Ome_res}
\end{figure}

Signatures of the DC resonances can be seen up to $\Omega \lesssim 1=v^*$, see Fig.~\ref{fig:fig4}(a)(b) and \ref{fig:fig6}(a)(b).
 Most prominently, one can observe a peak around $E_0\simeq U/2$, which becomes less prominent with increasing $\Omega$, but remains visible as a hump.
 This is consistent with our analysis of the spectral functions in Fig.~\ref{fig:fig3s}, which showed that in this driving regime, 
 the WS peaks still show up in the local spectrum. 
The position of this hump tends to gradually shift to larger values of $E_0$ with increasing $\Omega$, and very roughly follows the line $U=2E_0-\Omega$, see the inset of Fig.~\ref{fig:fig6}(a). 
(One can better see the tendency in $d_0$.)
While this could be an evidence for combined tunneling and photon emission a simpler explanation of the shift is that in the AC regime, the average field strength is reduced, so that a larger $E_0$ is needed for a significant doublon-holon production. 
By further increasing the excitation frequency, we enter the crossover regime, where the current and double occupancy show very nontrivial dependence 
both on $\Omega$ and $E_0$, see Fig.~\ref{fig:fig6}(c)(d).

 \begin{figure}[t]
  \centering
    \hspace{-0.cm}
    \vspace{0.0cm}
   \includegraphics[width=86mm]{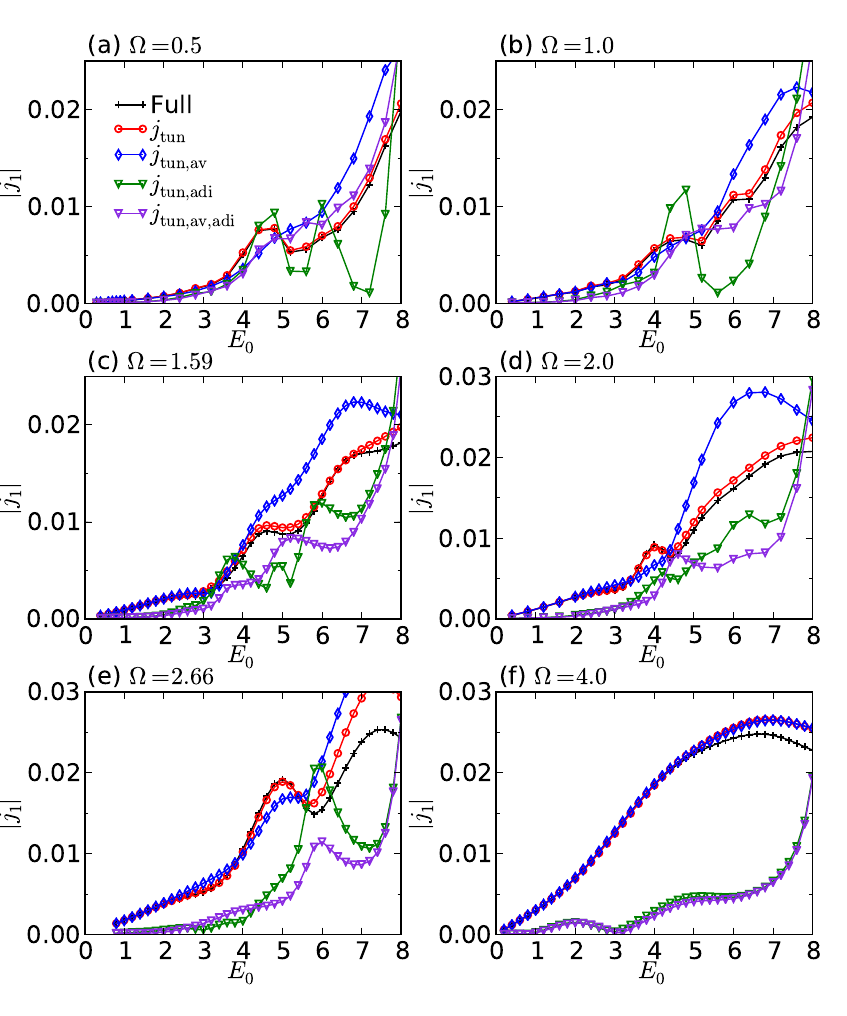} 
  \caption{Comparison of $|j_1|$ evaluated by different tunneling formulas for selected excitation frequencies 
  near the resonances $n\Omega=U$. Here, $U=8.0,\beta=2.0,\Gamma=0.06,W_{\rm bath}=5.0$.}
  \label{fig:fig8}
\end{figure}

At a first glance, the lines $U=E_0+\Omega$, $U=E_0+2\Omega$ and $J_0(E/\Omega)=0$ also appear to be correlated with the structures revealed by the numerical simulations.
However, it is difficult to obtain conclusive evidence of processes involving tunneling between the Floquet sidebands. 
 As an example, 
 let us take a closer look at the $U=2E_0+\Omega$ line, which seems to correlate with an increase in $d_{0}$ and $j_1$. In Fig.~\ref{fig:fig_U_2E_Ome_res} we plot the double occupation and the current (almost) at the resonances $\Omega=U/n$ as a function of $E_0$. 
The upper panels show the original data, while the lower panels plot these curves as a function of $E_0-E_{0,\text{line}}$, where $E_{0,\text{line}}=\frac{U-\Omega}{2}$ is the ``threshold value" predicted by the line $U=2E_0+\Omega$. This rescaling of the field strength roughly collapses the curves for the double occupation and the current. In particular, the upturn in the double occupation and current is roughly explained by $E_{0,\text{line}}$, which may suggest an interplay between field-induced tunneling and photon absorption. 

We can check to what extent the concept of tunneling between Floquet sidebands is meaningful by considering the tunneling formula for the current, and its simplified versions. 
In particular, the approximation $j_{\rm tun,av,adi}$ defined in Eq.~(\ref{eq:j_tun_av_adi}) is based on a picture of field-induced tunneling between the Floquet sidebands of the time-averaged spectral function.

Figure~\ref{fig:fig8} compares the $E_0$-dependence of the current near the resonances $\Omega=U/n$ to the prediction of the tunneling formula and its approximations. 
First, we note that the full tunneling formula Eq.~(\ref{eq:j_tun_tot}) is quantitatively accurate as long as the field is not too strong, and it is qualitatively good enough to capture the characteristic features of $|j_1|$.
The simplified version Eq.~(\ref{eq:j_tun_av}) ($j_\text{tun,av}$) based on $t_\text{av}$-averaged spectral functions can qualitatively reproduce almost all the characteristic 
structures except for the peak at $E_0=4$ and $\Omega=2.0$.
As expected, it becomes close to $j_\text{tun}$ as $\Omega$ becomes larger.
The current  defined in Eq.~(\ref{eq:j_tun_adi}) ($j_\text{tun,adi}$), which is calculated under the assumption of adiabatically varying fields, is generally less accurate than $j_\text{tun}$ and $j_\text{tun,av}$.
It can capture the behavior around $E_0=4=U/2$ for small $\Omega$, but it generally predicts complicated and unphysical structures.
Still we note that for $\Omega=0.5$, it is semi-qualitatively good up to $E_0=5$ as expected from the adiabatic condition.
The current  $j_\text{tun,av,adi}$ (Eq.~(\ref{eq:j_tun_av_adi})), which is calculated from $t_\text{av}$-averaged spectral functions under the assumption of adiabatically varying fields, is less accurate than $j_\text{tun}$ and $j_\text{tun,av}$. 
It also captures the peak structures around $E_0=4=U/2$ at small $\Omega$, but shows a complicated and strong dependence on the field.
 We note that $j_\text{tun,av,adi}$ does predict an upturn at roughly the expected value of $E_{0,{\rm line}}=(U-\Omega)/2$ (except for $\Omega=4$), but significantly underestimates the current in the weak-field regime for $\Omega\gtrsim 1$. We thus conclude that while combined tunneling and multi-photon absorption processes may contribute to the structures in Fig.~\ref{fig:fig4}, a simple picture based on time averaged Floquet spectral functions cannot capture the full complexity of the nonequilibrium processes that govern the generic AC driven system.

 \begin{figure}[t]
  \centering
    \hspace{-0.cm}
    \vspace{0.0cm}
   \includegraphics[width=86mm]{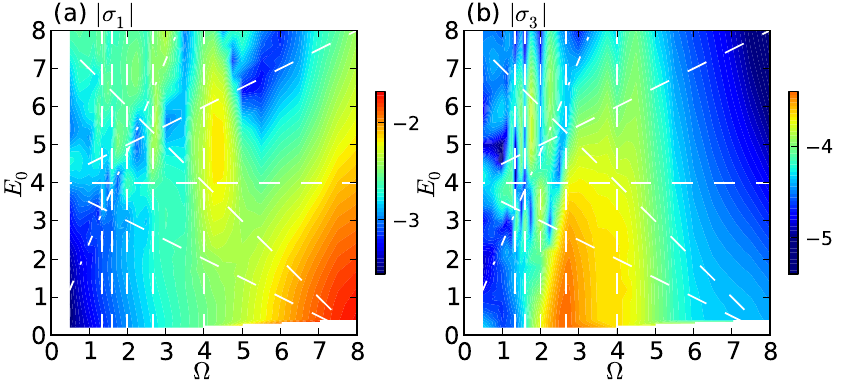} 
  \caption{(a)(b)  Log-scale plot of the linear ($|\sigma_1|$) and third harmonic ($|\sigma_3|$) conductivity. Here, $U=8.0,\beta=2.0,\Gamma=0.06,W_{\rm bath}=5.0$. }
  \label{fig:fig_optcond}
\end{figure}

In order to see how well the induced current is described by a perturbative process,
we evaluate the optical conductivities as 
\begin{align}
\sigma_n = j_n/E_0^{|n|}.
\end{align}
The lowest order contribution for the $n$-th component current can be regarded as a process where the system 
absorbs $n$ photons and generates a $n\Omega$ response, which scales as $E_0^{|n|}$.
In Fig.~\ref{fig:fig_optcond}, we show the results for $n=1$ and $3$.
For $n=1$, the conductivity is large around $U=8$ in the weak-field regime.
This is identical to the peak in the optical conductivity of the Mott state around $\omega=U$, which corresponds to the one photon absorption process.
For $n=3$, there is a strong signal around $\Omega=U/3$ in the weak-field regime, which corresponds to the three photon absorption process between Hubbard bands.
As we increase the field strength, the system goes into the nonperturbative regime, and $\sigma_n$ starts to exhibit a non-monotonic behavior.
One can identify peak structures at the resonances ($\Omega=U/n$) or between them, but in general it is difficult to identify the origin of these structures.
Higher harmonic components in the current are related to high-harmonic generation (HHG), which we discuss in the next section.

\subsection{High-harmonic generation and tunneling formula}
\label{sec:hh_and_tunneling}

In this section, we discuss how well the tunneling formulas can reproduce the higher order components 
of the current and hence the high-harmonic generation (HHG) in Mott insulators.
HHG originates from the strong interaction between light and matter.\cite{Corkum1993,Lewenstein1994,Cavalieri2007,Krausz2009} 
Recently, it has been observed in semiconductors and the possibility of HHG in solids is attracting interest.\cite{Golde2008,Ghimire2010,Schubert2014,Hohenleutner2015,Luu2015,Kemper2013b,Higuchi2014,Vampa2014,Vampa2015,Vampa2015b,Langer2016,Ndabashimiye2016,Liu2016,You2016,Otobe2016,Luu2016,Tamaya2016,Yoshikawa2017,Ikemachi2017,Tancogne-Dejean2017b,Tancogne-Dejean2017} 
While most of the studies so far have focused on semiconductors, there has been some recent effort to extend this concept 
to strongly correlated systems.\cite{Ivanov2017,Tancogne-Dejean2017c,Murakami2017d}

 \begin{figure}[t]
  \centering
    \hspace{-0.cm}
    \vspace{0.0cm}
   \includegraphics[width=70mm]{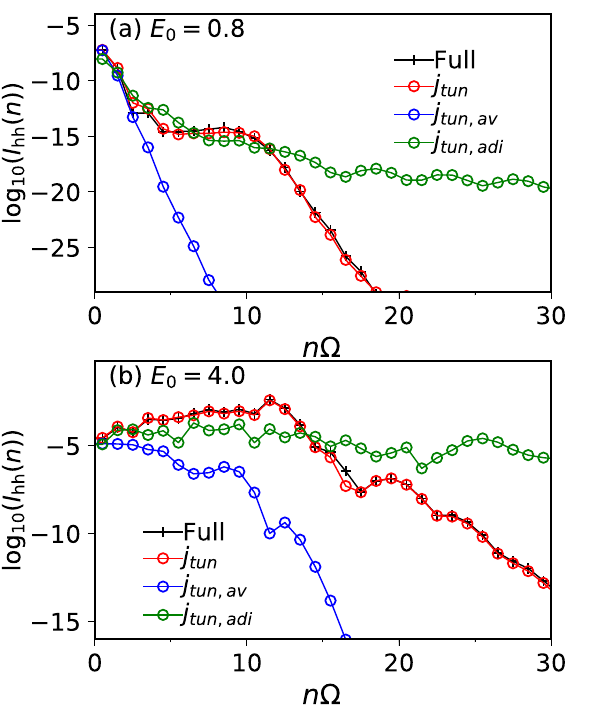} 
  \caption{Comparison of the high-harmonic generation spectra evaluated by different tunneling formulas. 
  Here, $U=8.0,\beta=2.0,\Gamma=0.06,W_{\rm bath}=5.0$.}
  \label{fig:fig7}
\end{figure}

In a previous study, Ref.~\onlinecite{Murakami2017d}, we investigated HHG in Mott insulators using FDMFT and 
the tunneling formula, which allowed to distinguish the contributions from recombination and hopping of doublons and/or holons.
Here we show complementary results which allow to assess the accuracy of the tunneling formula and its simplified versions. In the present case, the HHG spectrum is given by 
$I_{\rm hh}(n\Omega)=|n\Omega j(n\Omega)|^2$ with  $n\in \mathbb{Z}$,~\cite{Kemper2013b,Tamaya2016,Ivanov2017}
which is proportional to the power radiated at the given frequency. 
This is because the acceleration of the charges generates electromagnetic field radiation.
 
 In Ref.~\onlinecite{Murakami2017d}, it has been pointed out that the creation mechanism of HHG in the Mott state 
 is different in the weak and strong field regimes.
 In Fig.~\ref{fig:fig7}, we show results representing these two regimes.
 In both regimes, the full tunneling formula Eq.~(\ref{eq:j_tun_tot}) almost perfectly reproduces the exact result.
 This indicates that Eq.~(\ref{eq:j_tun_tot}) is powerful enough to correctly predict the high-order components of the current.
 On the other hand, the simplified formulas based on the time-averaged Green's functions ($j_{\rm tun,av}$,  Eq.~(\ref{eq:j_tun_av})) and the adiabaticity of the field ($j_{\rm tun,adi}$, Eq.~(\ref{eq:j_tun_adi})) 
 fail to reproduce the characteristic plateau structures.
 To be more precise, in the weak-field case, both $j_{\rm tun,av}$ and $j_{\rm tun,adi}$ roughly reproduce the results for the  low-order harmonics 
 in the weak-field regime, while 
 in the strong field regime $j_{\rm tun,adi}$ roughly reproduces the results there.
 On the other hand, both formulas fail to reproduce the higher-order harmonics, including the plateaus. 
 These results show that the HHG is the consequence of a non-trivial interplay between the non-adiabaticity of the field 
 and the dependence of the Green's function on the average time.

\section{Conclusions}\label{sec:conclusions}
We systematically investigated the properties of the NESSs realized in large-gap Mott insulators driven by DC or AC electric fields. 
To this end, we applied the Floquet DMFT (implemented with an NCA impurity solver) to the Hubbard model coupled to a free electron bath.
We also introduced a generalized tunneling formula (Eq.~(\ref{eq:j_tun_tot})) for the current, which resembles the Meir-Wingreen formula, 
as well as simplified versions neglecting the $t_\text{av}$-dependence of the Green's functions, or assuming an adiabatic evolution of the field. 
The full formula is derived by considering the tunneling between two neighboring slabs in Floquet steady states,
and it provides a direct relation between the structures of the single particle Green's function and the induced current.
We also derived the exact expression of the current in DMFT (Eq.~(\ref{eq:j_exact}))
and pointed out the relations between the tunneling formulas.

First, we considered DC driven systems and demonstrated the emergence of Wannier-Stark peaks from the lower and upper Hubbard bands in the spectrum. This result is consistent with the previous literature.\cite{Freericks2008FK,Eckstein2013b,Werner2015} 
In addition, we revealed WS peaks originating from many-body processes, which are absent in semiconductors.
These states can be connected to peaks and humps in the induced DC current through the tunneling formula.
Second, we studied the generic AC driven system and thus shed light on the complicated crossover regime which
connects the DC limit and the weak-field AC limit.

When the excitation frequency $\Omega$ is small enough ($\Omega \lesssim v^*$), one can still identify the WS peaks in the time-averaged spectrum and the behavior of the current and double occupancy resembles the result for the DC case. 
In the crossover regime, i.e. when $v^*\lesssim \Omega,E_0$, the behavior of the current and the double occupation is in general very complicated. 
Characteristic resonance structures at $\Omega=U/n$ in the double occupation and the current can be observed when the field is strong enough but not too strong ($E_0\lesssim U/2$). 
When the field becomes stronger ($E_0\gtrsim U/2$), the higher-$n$ resonances start to be affected by tunneling processes, which leads to the suppression of peaks or shifts to non-resonant frequencies. 
Rough data collapses along certain lines (e. g. $U=2E_0+\Omega$) suggest a possible cooperative effect between tunneling and photon-absorption, although an analysis based on the simplified tunneling formulas showed that the naive picture of tunneling between sidebands of the time-averaged Floquet spectra cannot fully capture the properties of the driven states. 
The full tunneling formula, on the other hand, works well in the generic AC case, and this may provide a basis for further analysis of the physical processes governing the crossover region. 

We have also studied the vicinity of the regime where the effective static Hamiltonian 
obtained from a high-frequency expansion becomes a free Hamiltonian.
In contrast to the Floquet prethermal states in isolated systems,\cite{Murakami2017c} the NESSs exhibit a time-averaged distribution function $\bar f(\omega)$ which is not periodic in $\omega$ (with period $\Omega$). This indicates that the steady state cannot be described as an equilibrium state of the effective static Hamiltonian. 
Still, reflecting the inverted population of Floquet prethermal states, one can identify driving regimes where the kinetic energy is positive.

Finally, we discussed the ability of the tunneling formula and its simplified versions to describe the HHG spectrum of Mott insulators.
The full tunneling formula is quantitatively good, while the simplified versions fail to capture the HHG features and the plateau structures in the HHG spectrum. 

The manipulation of systems by continuous periodic excitations (so-called Floquet engineering\cite{Oka2009,Jotzu2014,Sentef2015Floquet,Mentink2015,Takasan2017,Oka2018,Gorg2018}) is a promising strategy for the exploration of new properties or functionalities of materials.
We hope that our systematic study of Floquet NESSs of Mott insulators, covering a wide range of excitation conditions, may serve as a reference point  
for further investigations of Floquet steady states in strongly correlated systems.
In addition, the generalized tunneling formula is useful to obtain insights into the processes that contribute to the current in Floquet steady states. 
It would be an interesting future work to extend this approach to more complicated models including multi-orbital and electron-phonon problems, in order to reveal the underlying physics.

\acknowledgements

We thank M. Eckstein and M. Sch\"uler for helpful discussion. This work was supported by the Swiss National Science Foundation through NCCR MARVEL and the European Research Council through ERC Consolidator Grant 724103. The calculations have been performed on the Beo04 cluster at the University of Fribourg, and the CSCS Dora cluster provided by MARVEL.

\appendix
\section{Derivation of the tunneling formula for NESSs}\label{app:tunneling}
Here we introduce the generalized tunneling formula for the current in the nonequilibrium steady states (NESS),
which helps us to understand physical processes involved. 
To derive the formula, we first select one direction in the hyper-cubic lattice ($x$) and 
regard the system as a stack of $(d-1)$-dimensional slabs, which are aligned in the $x$ direction.
The Hamiltonian can now be expressed as $\hat{H}(t)=\hat{H}_{\perp}(t)+\hat{V}_{x}(t)$ with 
\small
\begin{align}
&\hat{H}_{\perp}(t)=-\sum_{\langle i,j\rangle_{\perp},\sigma} v_{ij}(t) c_{i,\sigma}^\dagger c_{j,\sigma}+U\sum_i n_{i\uparrow}n_{i\downarrow}+H_{\rm bath},\\
&\hat{V}_{x}(t)=-\sum_{\langle i,j\rangle_\parallel,\sigma} v_{ij}(t) c_{i,\sigma}^\dagger c_{j,\sigma}.
\end{align}
\normalsize
Here $\langle i,j\rangle_{\perp}$ indicates ${\bf r}_i-{\bf r}_j \perp {\bf e}_x$ and $\langle i,j\rangle_{\parallel}$ indicates ${\bf r}_i-{\bf r}_j \parallel {\bf e}_x$.
In the Floquet steady state, the initial correlations should be washed out because of the heat bath.
Therefore, one can prepare the steady states of the full system by starting from a steady state of $\hat{H}_{\perp}$, where all slabs are disconnected, and adiabatically switching on $\hat{V}_{x}$.
When $U$ is large, the effect of $\hat{V}_{x}$ may be treated perturbatively.
Here we consider the linear contribution and evaluate the current in the $x$ direction.

\begin{figure}[b]
  \centering
    \hspace{-0.cm}
    \vspace{0.0cm}
   \includegraphics[width=50mm]{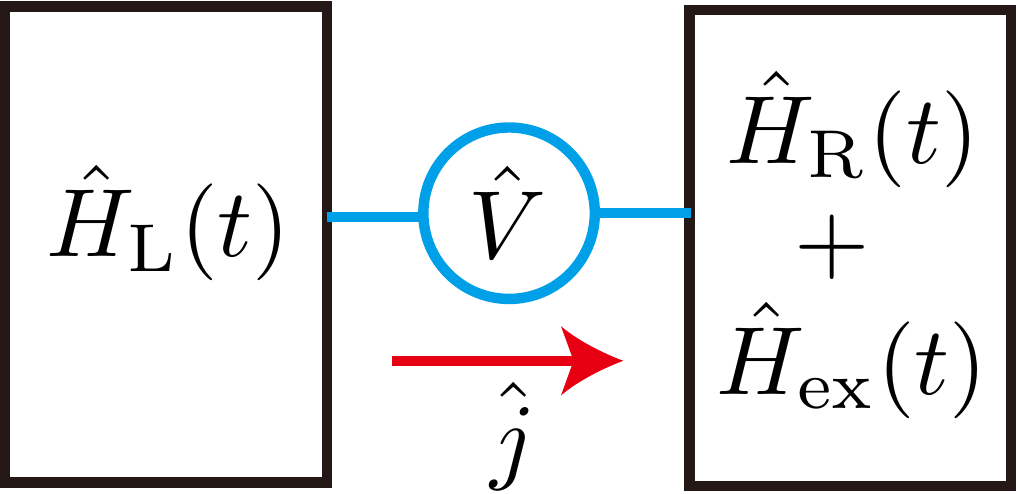} 
  \caption{Schematic picture of two subsystems in Floquet steady states, which are connected by junctions ($\hat{V}$).}
  \label{fig:Floquet_tunnel}
\end{figure}

In this setup, in order to evaluate the current that flows between a pair of neighboring slabs, we only need to focus on these two neighboring slabs, 
because processes involving several slabs are coming from higher oder processes in $\hat{V}_{x}$.
We denote these slabs by L and R.
After changing the gauge in the $x$ direction to the scalar gauge, the bias is applied to the $x$ direction and each slab is connected to the neighboring slab by $l^{d-1}$ 
tunneling junctions. Here $d$ is the spatial dimension, we consider $l^d$ lattices and take the thermodynamic limit $l\rightarrow \infty$ in the end.
The total Hamiltonian of these two slabs and the junctions between them can be written as 
\small
$\hat{H}_{\rm tot}(t)=\hat{H}_{\rm tot,0}(t)+\hat{V}=\hat{H}_{\rm L}(t)+\hat{H}_{\rm R}(t)+\hat{H}_{\rm ex}(t)+\hat{V}$ with
\begin{align}
\hat{H}_{\rm L}(t)&=\hat{H}_{\rm L}(t;{c^\dagger_{L,i_\perp,\sigma},c_{L,i_\perp,\sigma}}),\nonumber\\
\hat{H}_{\rm R}(t)&=\hat{H}_{\rm R}(t;{c^\dagger_{R,i_\perp,\sigma},c_{R,i_\perp,\sigma}}),\nonumber\\
\hat{H}_{\rm ex}(t)&=E(t)\sum_{i_\perp,\sigma} n_{R,i_\perp,\sigma},\\
\hat{V}&=-v\sum_{\sigma,i_\perp}[c_{\rm L,i_\perp,\sigma}^\dagger c_{R,i_\perp,\sigma}+c_{\rm R,i_\perp,\sigma}^\dagger c_{L,i_\perp,\sigma}].\nonumber
\end{align}
\normalsize
In Fig.~\ref{fig:Floquet_tunnel}(a), we illustrate this situation.
Here $i_\perp$ is the site index perpendicular to $x$, and $\hat{H}_{\rm L,R}$ conserves the number of particles in the subsystems.
We note that L and R are decoupled initially and the system is  described by the time periodic Hamiltonian $\hat{H}_{\rm tot,0}(t)$.  
Under the continuous driving each slab is in a time-periodic steady state (Floquet steady state).
Hence the initial wave function is one of the Floquet solutions of $\hat{H}_{\rm tot,0}(t)$,
\small
\begin{align}
|\Psi^{(0)}_{\rm tot,\alpha\beta}(t)\rangle_s = |\Psi^{(0)}_{\rm L,\alpha}(t)\rangle_s \otimes  |\Psi^{(0)}_{\rm R,\beta}(t)\rangle_s e^{-i\int^t d\bar{t} E(\bar{t})N_{R,\beta}}.
\end{align}
\normalsize
Here the subscript ``$s$" indicates the Schr\"odinger representation and $|\Psi^{(0)}_{\rm X,\alpha}(t)\rangle_s$ is the Floquet solution of the time periodic Hamiltonian $\hat{H}_{\rm X}$ (X=R,L), 
whose indices are $\alpha$. $N_{\rm R,\beta}$ is the number of particles in the right subsystem.
We adiabatically switch on $\hat{V}$ 
from a certain time $t_0$ and wait for a long time and measure the induced current between the subsystems,
\begin{align}\label{eq:j_ope}
\hat{j}=iqv\sum_{\sigma,i_\perp,\sigma}[c_{\rm L,i_\perp,\sigma}^\dagger c_{R,i_\perp,\sigma}-c_{\rm R,i_\perp,\sigma}^\dagger c_{L,i_\perp,\sigma}].
\end{align}
The first order correction of $\hat{V}$ to the state is 
\begin{align}
|\Psi_{\rm tot,\alpha\beta}(t)\rangle_s\simeq|\Psi^{(0)}_{\rm tot,\alpha\beta}(t)\rangle_s-i\int^t_{-\infty}d\bar{t}\hat{\mathcal{U}}_0(t,\bar{t}) \hat{V} |\Psi^{(0)}_{\rm tot,\alpha\beta}(\bar{t})\rangle_s.
\end{align} 
Here $\hat{\mathcal{U}}_0(t,t')=\mathcal{T}\exp[-i\int^t_{t'} d\bar{t} \hat{H}_{\rm tot,0}(\bar{t})]$ for $t>t'$
and $\mathcal{T}$ is the normal time-ordering operator.
From this we obtain the current per junction,
\begin{align}
j(t)=-i\frac{1}{l^{d-1}}\int^t_{-\infty}d\bar{t}s\langle \Psi^{(0)}_{\rm tot,\alpha\beta}(t) | \hat{j}\hat{\mathcal{U}}_0(t,\bar{t}) \hat{V} |\Psi^{(0)}_{\rm tot,\alpha\beta}(\bar{t})\rangle +\text{H.c.}.
\end{align}
Here  the zero-th order term vanishes because of particle conservation. 
Using the fact that $|\Psi^{(0)}_{\rm tot,\alpha\beta}(t)\rangle_s$ and $\hat{\mathcal{U}}_0(t,t')$ is the direct product of components of the left and right subsystems,
we can derive the following form of the current,
\small
\begin{align}\label{eq:j_tun_orig}
&j(t)=-2qv^2 \frac{1}{l^{d-1}}\sum_{i_\perp,j_\perp}\int^t_{-\infty} d\bar{t} 
\bigl\{G^<_{\rm L,\alpha,i_\perp j_\perp}(\bar{t},t) G^>_{\rm R,\beta,j_\perp i_\perp}(t,\bar{t})\nonumber\\
&-G^>_{\rm L,\alpha,i_\perp j_\perp}(\bar{t},t) G^<_{\rm R,\beta,j_\perp i_\perp}(t,\bar{t}) \bigl\}
e^{-i\int^{t}_{\bar{t}}dt'E(t')}+\text{H.c.}.
\end{align} 
\normalsize
Here the factor of two comes from the spin degrees of freedom and for $X=L,R$ and $\gamma=\alpha,\beta$,
\begin{subequations}
\begin{align}
G^<_{X,\gamma,i_\perp j_\perp}(t',t)&=i\langle \Psi^{(0)}_{\rm X,\gamma}(t) |c^\dagger_{X,j_\perp}\mathcal{U}_{0X}(t,t') c_{\rm X,i_\perp}| \Psi^{(0)}_{\rm X,\gamma}(t') \rangle,\\
G^>_{X,\gamma,i_\perp j_\perp}(t,t')&=-i\langle \Psi^{(0)}_{\rm X,\gamma}(t) |c_{X,i_\perp}\mathcal{U}_{0X}(t,t') c^\dagger_{\rm X,j_\perp}| \Psi^{(0)}_{\rm X,\gamma}(t') \rangle.
\end{align}\label{eq:G_LR}
\end{subequations}
So far we have assumed that initially the left and right subsystems are in certain Floquet states, respectively.
In reality the Floquet steady state should be a mixed state of all Floquet states,
$\rho_{\rm X}(t)=\sum_{\gamma} \rho_{\gamma} |\Psi^{(0)}_{\rm X,\gamma}(t)\rangle \langle \Psi^{(0)}_{\rm X,\gamma}(t)|$,
and we should use this as the initial state. Here $\rho_{\gamma}$ is some weight factor.  
The expression of the current is obtained by taking the average over all states.
Then we can remove the index $\alpha,\beta$ in Eq.~(\ref{eq:j_tun_orig}) and 
the definition of the Green's functions becomes Eq.~(\ref{eq:G_LR}) averaged over different Floquet states.
In the present case, the time-periodic Hamiltonian is the same for the left and right systems and we can write $G_{\rm R}=G_{\rm L}=G^{'}$.
Here we put a prime to emphasize that $G^{'}$ is the Green's function for the $(d-1)$-dimensional slab.
This leads to the expression 
\begin{align}\label{eq:j_tun_full2}
j(t)=&-2qv^2 \frac{1}{l^{d-1}}\sum_{i_\perp,j_\perp}\int^t_{-\infty} d\bar{t} 
\bigl\{G^{'<}_{i_\perp j_\perp}(\bar{t},t) G^{'>}_{j_\perp i_\perp}(t,\bar{t})\nonumber\\
&-G^{'>}_{i_\perp j_\perp}(\bar{t},t) G^{'<}_{j_\perp i_\perp}(t,\bar{t}) \bigl\}
e^{-i\int^{t}_{\bar{t}}dt'E(t')}+\text{H.c.}.
\end{align} 
On the other hand, since we are in the large $d$ limit, $G^{'}$ can be replaced by $G$ in
the full $d$-dimensional bulk. Expressing Eq.~(\ref{eq:j_tun_full2}) in terms of the momentum in the $(d-1)$-dimensional space 
and replacing $G^{'}$ with $G$, we obtain Eq.~(\ref{eq:j_tun_mom}).

Now we further simplify Eq.~(\ref{eq:j_tun_full2}) by only considering the process where $i_\perp=j_\perp$.
Physically this represents the contribution to the current at a certain junction between L and R by an electron that 
hopped through the same junction. 
Finally taking account of $v=\frac{v^*}{2\sqrt{d}}$ and the contribution from all directions,
we obtain Eq.~(\ref{eq:j_tun_tot}).

\section{Semiconductor model}\label{app:semi}

In order to clarify the effects associated with correlations, we consider a semiconductor model described by 
\begin{align}
H_{\rm semi}(t)
=-&\sum_{\langle i,j\rangle,\alpha}v^\alpha_{ij}(t) c_{i\alpha}^\dagger c_{j\alpha}
-\sum_{\langle i,j\rangle}v^{cv}_{ij}(t) (c^\dagger_{ic}c_{jv}+c^\dagger_{iv}c_{jc})\nonumber\\
&+\sum_{i,\alpha}D_{\alpha}c^\dagger_{i\alpha}c_{i\alpha}.\label{eq:H_semicon}
\end{align}
Here $\alpha=c,v$ is the orbital index and we consider one valence band and one conduction band,
which correspond to the lower and upper Hubbard bands, respectively. $D_c=-D_v=U/2$ is the crystal field.
In order to mimic the fact that, in the Hubbard model, doublons and holons are created at neighboring sites by the hopping of an electron,
we consider a transfer integral between neighboring sites in the semiconductor model.
The effect of the field is taken into account through the Peierls substitution. 
We consider the NESS of the system coupled to a free electron bath, 
choose $v^c=-v^{v}=v^{vc}=0.5v$ and a hyper-cubic lattice.

 \begin{figure}[t]
  \centering
    \hspace{-0.cm}
    \vspace{0.0cm}
   \includegraphics[width=86mm]{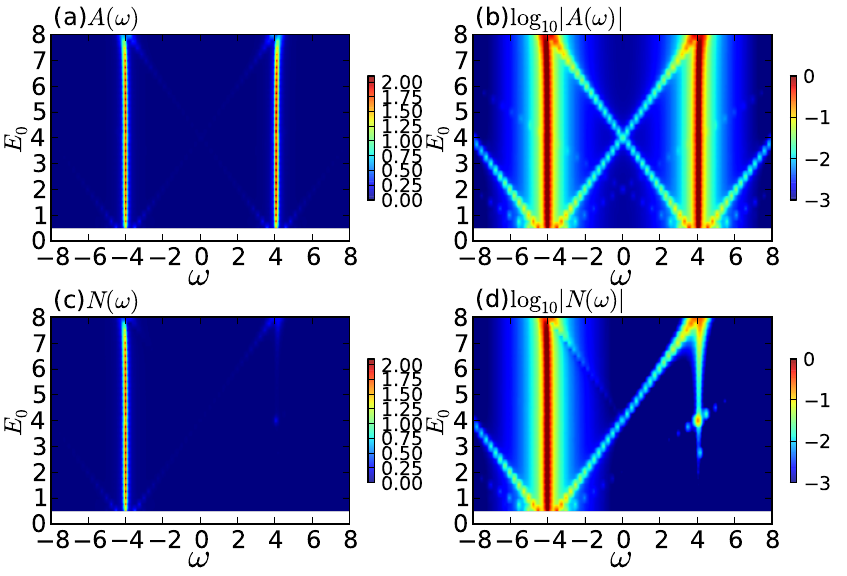} 
  \caption{(a)(b) Local spectral function ($A(\omega)$) and (c)(d) occupation function ($N(\omega)$) of the semiconductor model with a DC field.  
  Here, $U=8.0,\beta=2.0,\Gamma=0.06$.}
  \label{fig:APX_fig1}
\end{figure}

 \begin{figure}[b]
  \centering
    \hspace{-0.cm}
    \vspace{0.0cm}
   \includegraphics[width=86mm]{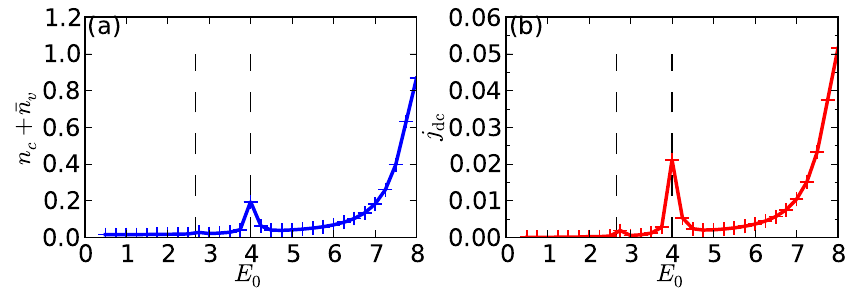} 
  \caption{(a)(b) Field-strength dependence of the number of charge carriers ($n_c+\bar{n}_v$), which corresponds to the double occupancy in the Mott insulators, and the current ($j_{\rm dc}$) of the semiconductor model with a DC field. 
  Vertical lines indicate $E_0=U$ and $U/2$. Here, $U=8.0,\beta=2.0,\Gamma=0.06$.}
  \label{fig:APS_fig2}
\end{figure}

In Fig.~\ref{fig:APX_fig1}, we show the spectrum ($A(\omega)=(A_{c}(\omega)+A_{v}(\omega))/2$) and occupation ($N(\omega)=(N_{c}(\omega)+N_{v}(\omega))/2$) for the DC field.
One can observe the Wannier-Stark peaks emanating from the valence and conduction bands.
On the other hand, there is no signature of the peak at $\omega=\pm\frac{3}{2} U\mp 2E$,
which demonstrates that many-body effects cause the appearance of this peak in the Mott insulating system.

In Fig.~\ref{fig:APS_fig2}, we show the number of charge carriers ($n_c+\bar{n}_v\equiv c_{i,c}^\dagger c_{i,c}+c_{i,v}c_{i,v}^\dagger$) 
and the induced DC current.
One can observe clear peaks at the resonances $\Omega=U/n$, as in the Mott insulator.
However, the structure at $\Omega=2U/3$ is missing, which is consistent with the absence of  $\omega=\pm\frac{3}{2} U\mp 2E$ sidebands
and supports the interpretation that this structure comes from many-body effects.

\section{Data Points}
\begin{figure}[t]
  \centering
    \hspace{-0.cm}
    \vspace{0.0cm}
   \includegraphics[width=50mm]{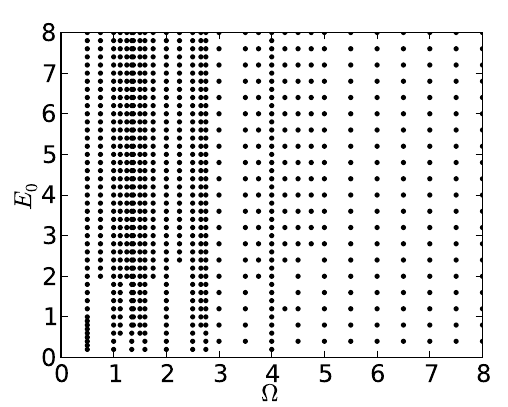} 
  \caption{Data points used for Fig.~\ref{fig:fig4} and Fig.~\ref{fig:fig_optcond}.}
  \label{fig:fig_data_points}
\end{figure}

Because of the strong dependence on frequency and field strength in the crossover regime, 
we explicitly show the data points used to draw Figs.~\ref{fig:fig4},\ref{fig:fig_optcond}.
Simulations were performed for the points shown in Fig.~\ref{fig:fig_data_points} and the remaining 
parameter values were obtained by linear interpolation.

\bibliography{Ref}

\end{document}